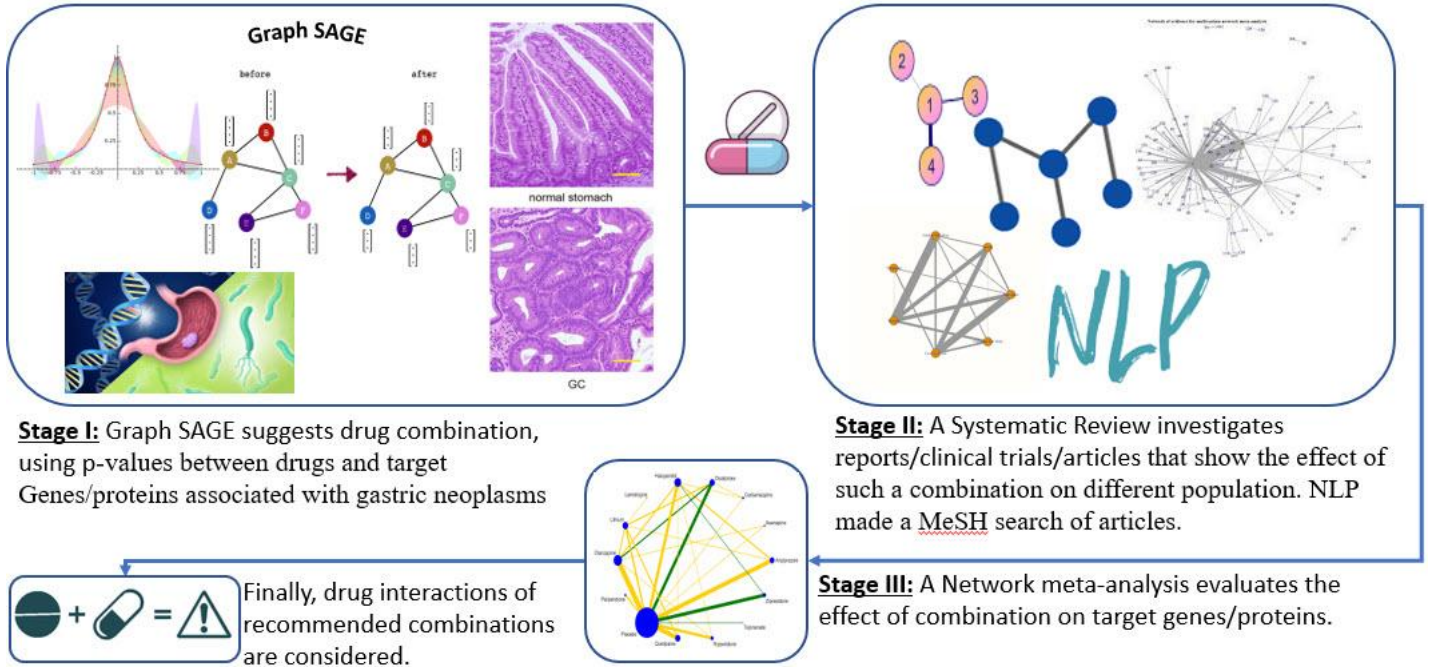

# Utilizing the RAIN method and Graph SAGE Model to Identify Effective Drug Combinations for Gastric Neoplasm Treatment

S. Z. Pirasteh[1], Ali A. Kiaei[2]*, Mahnaz Bush[3], Sabra Moghadam[4], Raha Aghaei[4], Behnaz Sadeghigol[5]

Highlights:
- Effective treatment strategies, particularly drug combinations, are essential for addressing the heterogeneity of the disease.
- The Graph SAGE model was utilized to propose potential drug combinations for treating gastric neoplasm.
- The proposed model recommended Fluorouracil, Trastuzumab, and Oxaliplatin as potential treatments for gastric cancer.
- Health policymakers can enhance patient care by utilizing the findings from artificial intelligence, specifically through the RAIN method.

[1] Department of Computer Engineering, Islamic Azad University South Branch, Tehran, Iran.
[2] Correspondence: ali.kiaei@sharif.edu Bioinformatics and Computational Biology (BCB) Lab, Sharif University of Technology, Tehran, Iran
[3] Cellular and Molecular Biology Research Center, Shahid Beheshti University of Medical Sciences, Tehran, Iran
[4] School of Medicine, Mazandaran University of Medical Science, Sari, Iran
[5] Department of Computer Engineering, Sharif University of Technology, Tehran, Iran

## Abstract:

**Background**: Abnormal cell growth in the stomach lining causes gastric neoplasm, also known as stomach cancer, which can affect any part of the stomach and has the potential to spread to other parts of the body. The most prevalent kind is adenocarcinoma. Since symptoms frequently appear only in advanced stages, resulting in complications like obstruction, bleeding, and metastasis, early detection and treatment are essential due to its aggressive nature and high mortality rate. In order to address the disease's heterogeneity, improve therapeutic outcomes, reduce resistance, and increase patient tolerance and quality of life, effective treatment strategies—especially drug combinations—are crucial.

**Method:** The RAIN method, which consists of three primary steps, was used in this investigation. Initially, potential pharmacological combinations for the management of gastric neoplasm were proposed utilizing the Graph SAGE model. The edges of the graph model are p-values that show the relationship between each node, which could be a drug, a human gene, or a protein that could be a target for gastric cancer. Second, studies detailing the recommended medications for the treatment of gastric neoplasms were identified through natural language processing (NLP) and a systematic review of the PubMed, Scopus, Web of Science, Embase, Science Direct, and Google Scholar databases. We then used network meta-analysis to compare how well each drug and its related genes worked. All implementations were done with Python software.

**Result**: the suggested model suggested oxaliplatin, fluorouracil, and trastuzumab as possible therapies for gastric cancer. The efficacy of these medications was confirmed by a review of 61 studies. A p-value of 0.0229 was found for the treatment impact of fluorouracil alone. The p-value increased to 0.0099 when fluorouracil and trastuzumab were taken together. The p-value further dropped to 0.0069 for the combination of all three medications, suggesting improved efficacy in treating gastric cancer.

**Conclusion**: health policymakers can improve patient care by applying the knowledge gained from artificial intelligence, particularly through the RAIN method, given the substantial impact of gastric neoplasm. This method can help choose the best medication combinations for managing and treating gastric neoplasms, providing a useful tactic to enhance patient outcomes.

# 1. INTRODUCTION:

Gastric neoplasms, also known as gastric cancer, are one of the most common types of cancer in the world. Their management and treatment have advanced significantly in recent years. The development of multimodality management strategies, for example, is changing clinical approaches by providing tailored therapies that target molecular vulnerabilities in both resectable and metastatic cases. This is demonstrated in studies such as Wen-Long Guan (2022).

Further evidence of continuous improvements in surgical techniques comes from developments like the confirmation of the long-term safety and effectiveness of laparoscopic distal gastrectomy in comparison to open surgery for locally advanced gastric cancer. Emerging research into perioperative immunotherapy and targeted therapies is filling critical gaps in treatment options for patients with advanced stages of the disease, as shown by trials like KEYNOTE-859 and SPOTLIGHT. Furthermore, the discovery of molecular biomarkers such as PD-L1, MSI, and HER2 presents encouraging opportunities for customizing treatments according to the unique features of each tumor. Despite these advancements, there are still issues, most notably the late diagnosis of many patients because there aren't enough distinguishing clinical indicators. To improve results and successfully handle the complexity of gastric neoplasms, more research is still necessary. [1]

For a number of strong reasons, treating gastric neoplasms, commonly referred to as gastric cancer, is extremely important. First off, patients with both surgically removable tumors and those with metastatic disease are increasingly receiving personalized medicine as the treatment landscape for gastric cancer rapidly changes due to the introduction of new therapies and surgical techniques. [2] Secondly, by combining immunotherapy, targeted therapies, and minimally invasive surgeries during the perioperative phase, these developments are completely changing the way that gastroesophageal cancer is treated. Furthermore, efficient treatment of gastric cancer is essential for both curing the illness and improving the prognosis of survivors. This includes things like nutritional support, post-treatment monitoring, and handling treatment-related side effects. Current research, like that done by Era Cobani (2024) highlight continuous attempts to improve treatment recommendations and maintain functional results without sacrificing cancer control in the Journal of Gastrointestinal Cancer. In summary, negotiating the challenges of treating gastric cancer is a dynamic field characterized by growing therapeutic alternatives and ongoing advancements in clinical procedures. [3]

The substantial benefits of using medication combinations to treat gastric neoplasms have been highlighted by recent studies. According to recent publications, these combinations have shown significant advantages:

For patients with metastatic colorectal cancer, combining therapies has significantly improved results. Patients treated with combination therapy had a median progression-free survival of 6.2 months versus 2.1 months and a median overall survival of 19.7 months versus 9.5 months when compared to those receiving targeted therapy alone.

In 17.3% of patients, the new combination treatments have resulted in partial or total tumor shrinkage.

Since zolbetuximab was first used in combination with traditional chemotherapy, patients with advanced gastric or gastroesophageal junction cancer who exhibit overexpression of the CLDN18.2 biomarker have had longer survival times.

Patients with advanced gastric cancer have shown improved survival rates and clinical benefits from immunotherapy advancements, such as the use of immune checkpoint inhibitors in conjunction with traditional treatments like chemotherapy and radiation therapy.

According to research by John T. Mullen (2024), patients with advanced gastric or gastroesophageal junction adenocarcinoma and CLDN18.2 protein overexpression may have a longer survival time if they receive the novel combination of zolbetuximab and conventional chemotherapy.[4]

These findings highlight how combined medication therapies have the potential to greatly increase the effectiveness of gastric neoplasm treatments, which would ultimately benefit patients.

## 1.1. Associated human genes/proteins:

CDH1 - One gene that is essential to gastric neoplasms is CDH1, also referred to as E-cadherin. It has long been believed that CDH1 is a tumor suppressor gene. Recent research, however, indicates that it might also act as a pro-oncogene, hastening the malignant development of specific cancers. A higher risk of breast and stomach cancers is linked to inherited variations in the CDH1 gene. As a result, a disorder called Hereditary Diffuse Gastric Cancer (HDGC) syndrome has been identified. Currently, prophylactic surgery and improved cancer surveillance techniques are used to manage this syndrome. Using organoids and animal models, researchers have discovered putative molecular drivers of HDGC development and are currently examining the effects of E-cadherin loss in the gastric epithelium. These findings hold promise for the development of biomarkers, targeted treatments, and chemoprevention tactics for diffuse-type gastric cancer. Because they have a lower cumulative risk than those with one or more family members who have gastric cancer, people with a CDH1 pathogenic or likely pathogenic variant who do not have a family history of the disease may benefit from gastric surveillance as an alternative to risk-reducing total gastrectomy. In summary, CDH1 functions as a tumor suppressor and a possible oncogene in gastric neoplasms. Future treatment approaches may target this gene since it plays a crucial role in the onset and spread of gastric cancer.[5,6]

FZR1 - The anaphase-promoting complex/cyclosome (APC/C) is a multi-subunit E3 ligase complex that needs FZR1, which is also known as Fizzy-Related Protein Homologue (Fzr), to work. It has been recognized as a possible target for treatment in multiple myeloma. FZR1 has been shown to be important in the context of multiple myeloma: Numerous myeloma primary cells and cell lines had elevated levels of FZR1. Multiple myeloma cell lines experienced growth arrest and decreased viability when FZR1 was knocked down. Topoisomerase IIα (TOPIIα), an APC/C Fzr substrate, accumulated as a result of FZR1 knockdown. The effects of treatment with proTAME, an inhibitor of both APC/C Fzr and APC/C Cdc20, were comparable. These findings suggest that FZR1 could serve as a promising therapeutic target in multiple myeloma; however, further research is necessary to determine its role and potential as a therapeutic target in gastric neoplasms. For the most accurate and up-to-date information, it is always best to talk to researchers or doctors. [7]

ERBB2 - One of the main therapeutic targets for gastroesophageal adenocarcinoma (GEA), a kind of gastric neoplasm, is ERBB2, also referred to as HER2. For 20 years, trastuzumab was the only treatment for GEA that overexpressed ERBB2. However, several medications that showed effectiveness in conjunction with or as alternatives to trastuzumab in breast cancer failed to exhibit clinical benefit in GEA. Because tumors are different from each other, there are low-expressing ERBB2 tumor clones, and ERBB2 levels are slowly dropping, GEA is more likely to be resistant to anti-ERBB2 therapy than other types of cancer. In order to combat heterogeneity in ERBB2-positive GEA, new techniques for ERBB2 testing and the application of antibody-drug conjugates with a bystander effect are being developed. Dual therapy can address co-amplifications of tyrosine kinase receptors, alterations in the MAPK and PI3K signaling pathways, and modifications in cell cycle-regulating proteins, all of which are recognized to play a role in resistance to anti-ERBB2 therapy. A plethora of evidence suggests that the efficacy of anti-ERBB2 therapy is predominantly facilitated by immune mechanisms, specifically antibody-dependent cell-mediated cytotoxicity, rather than intracellular signaling pathways. - In conclusion, ERBB2 is a prospective target for forthcoming therapeutic strategies due to its essential role in the onset and dissemination of gastric neoplasms. Nevertheless, tumor heterogeneity-induced resistance to anti-ERBB2 therapy presents a challenge that necessitates focus in ongoing and future research.[8,9]

MUC6 - In the study of cancer, MUC6, a form of mucin present in the stomach glands, has shown great promise. It works as a possible target in the following ways:Studies show that MUC6 expression is a useful prognostic indicator, especially in lung invasive mucinous adenocarcinoma. Premalignant lesions of the stomach, pancreas, and bile duct frequently exhibit decreased αGlcNAc glycosylation on MUC6-positive tumor cells, and invasive cancers in these organs are correlated with decreased MUC6 expression. Studies conducted in lab settings have shown that, in comparison to control cells, overexpression of MUC6 in A549 cells—which are derived from lung cancer with a KRAS mutation—significantly suppresses proliferation, motility, and invasiveness. Furthermore, it has been demonstrated that MUC6 reduces tumor aggressiveness by degrading β-catenin via autophagy. One important biomarker for early-stage epithelial neoplasms in the biliary tract, pancreas, uterine cervix, and stomach is thought to be decreased αGlcNAc glycosylation on the MUC6 scaffold. Even though these studies show that MUC6 may be a therapeutic target for a number of cancers, including gastric neoplasms, more research is necessary to completely understand its therapeutic potential and function in clinical settings. For the most up-to-date and accurate information on this subject, it is advised to speak with researchers or medical professionals.[10]

GAST - The hormone known as GAST, or gastrin, is necessary for the stimulation of the production of gastric acid, which is necessary for efficient digestion. Its clinical significance is highlighted by its involvement in a number of gastrointestinal disorders, most notably gastric neoplasms. Its potential as a therapeutic target is highlighted by recent studies: Gastrin's role in gastric cancer is pivotal, influencing the growth of gastric epithelial cells and implicated in the progression to adenocarcinoma. Moreover, it stimulates gastric acid secretion by parietal cells, crucial for maintaining the gastric mucosal barrier. Disruption of this barrier can lead to gastric ulcers, fostering conditions conducive to gastric cancer development. Additionally, gastrin's involvement in promoting gastric inflammation further complicates its role in disease progression. By enhancing the proliferation of gastric epithelial cells, gastrin contributes directly to the pathogenesis of gastric neoplasms. Targeting gastrin or its receptor presents a promising strategy for therapeutic intervention in the treatment of gastric neoplasms. Despite these insights, ongoing research is essential to elucidate gastrin's precise therapeutic potential and its implications for clinical practice. For the most accurate and current understanding, consulting healthcare professionals or researchers is recommended.[11,12]

TP53 - The CDKN2A/MDM2/p53 pathway controls cell cycle progression and DNA damage responses essential for tumor growth in gastrointestinal stromal tumors (GIST), and TP53, also known as p53, is a critical tumor suppressor gene implicated in a number of cancers, including gastric neoplasms. Recent research emphasises TP53's potential as a therapeutic target in the following ways: TP53 is a biomarker and therapeutic target in GIST, where it functions as apoptosis, autophagy, pyroptosis, ferroptosis, and apoptosis[2]. In contrast to mutant p53, which causes PCD, wild-type p53 enhances apoptosis-related proteins to promote apoptosis and preserve organism stability. Additionally, in patients with gastric cancer, TP53 mutations frequently coexist with other genetic changes, providing information about the course of the disease and possible treatment options. While TP53 presents promising opportunities for targeted therapies in gastric neoplasms, further research is necessary to fully delineate its therapeutic potential. For the most accurate and current understanding, consulting healthcare professionals or researchers is advised.[13,14]

CLDN18 - Claudin-18.2, another name for CLDN18, is a tight-junction protein that has been found to be a promising therapeutic target in gastric cancer. A monoclonal antibody called zolbetuximab (IMAB362), which targets CLDN18.2, is currently under investigation for the treatment of advanced gastrointestinal malignancies. Clinical trials, like the phase II FAST study, have demonstrated that zolbetuximab in conjunction with first-line EOX therapy improves overall survival (OS) and progression-free survival (PFS) in patients with advanced gastric adenocarcinoma who express CLDN18.2. With a strong correlation to Borrmann type, degree of differentiation, PD-L1 expression, and patient survival, CLDN18.2 expression is a useful prognostic indicator in gastric cancer. It has been identified as an independent risk factor affecting prognosis. CLDN18.2 plays a crucial role in facilitating adhesion between gastric cancer cells and cancer-associated fibroblasts (CAFs), promoting the formation of emboli and enhancing metastatic progression. CLDN18.2 expression is associated with PD-L1 expression in tumors and immune infiltration, specifically with higher fractions of CD4+ and CD8+ T cells. In conclusion, CLDN18.2 expression levels in gastric cancer tissues correlate with prognosis and play pivotal roles in tumor progression and the immune microenvironment. These insights highlight CLDN18.2 as a potential therapeutic target and underscore its role in promoting gastric cancer progression and metastasis through interactions with CAFs[2]. Further research is essential to fully exploit CLDN18.2-targeted therapies in clinical practice. For the most current and precise information, consultation with healthcare professionals or researchers is recommended.[15,16]

GKN1 - Gastrokine 1, or GKN1, is a protein that is mostly found in the stomach and has drawn interest as a possible treatment target for gastric cancer. As an autocrine/paracrine protein, GKN1 inhibits cell proliferation by triggering senescence pathways that involve p16/Rb and p21 waf. Research using animal models and gastric cancer cells has shown that GKN1 treatment can reduce persistent Ras/Raf/MEK/ERK signaling, which is linked to the development of cancer. One potential diagnostic marker for gastric cancer is GKN1. Numerous mammalian species and healthy individuals have had it found in their stomach mucosa cells, indicating its possible use in early detection techniques. Although the exact molecular role of GKN1 in gastric cancer is still unknown, systematic research is being done to clarify its function and potential therapeutic applications in cancer cells. In conclusion, GKN1 is a key player in the pathophysiology of gastric neoplasms and may be a target for treatment. To completely comprehend its mechanisms and evaluate its therapeutic potential in clinical applications, more research is necessary. It is recommended to speak with researchers or medical professionals for the most recent information.[17,18]

CDX2 - A transcription factor essential for intestinal development and differentiation, CDX2 (Caudal Type Homeobox 2) has been found to be a promising therapeutic target in gastric neoplasms. With positive expression strongly correlated with better differentiation and a better prognosis for patients, CDX2 exhibits promise as a prognostic marker in gastric carcinoma. In human gastric carcinomas, CDX2 expression tends to gradually decline. CDX2 transgenic mice have been used in studies to show that it can change gastric mucosa into intestinal metaplastic mucosa. In some models, it has also been shown to progress to gastric cancer. Many gastric cancers and intestinal metaplasia (IM) express CDX2, suggesting its involvement in gastric cancer development. To sum up, CDX2 is a key player in the development and spread of gastric neoplasms, making it a viable target for treatment. To completely understand its mechanisms and assess its therapeutic potential in clinical settings, more research is necessary.[19]

MUC5AC - With implications for many cancer types, the gastric gland mucin MUC5AC has become a promising target in cancer research. Expression of MUC5AC is a useful prognostic indicator, especially in cases of lung invasive mucinous adenocarcinoma. Premalignant lesions of the stomach, pancreas, and bile ducts have been found to have altered αGlcNAc glycosylation on MUC5AC-positive tumor cells; decreased expression of this glycosylation is linked to invasive cancer in these organs. Experimental research indicates that ectopic MUC5AC expression in colon cancer cells inhibits their capacity to form colonies by reducing cell viability, promoting apoptosis, and inducing G1 cell cycle arrest. One important biomarker for early-stage gastric, pancreatic, biliary tract, and uterine cervix epithelial neoplasms is decreased αGlcNAc glycosylation on MUC5AC. MUC5AC may be a therapeutic target for a variety of cancer types, according to these findings, but further investigation is required to fully elucidate its mechanism and therapeutic efficacy, particularly in gastric neoplasms.[20]

RUNX3 - Runt-related transcription factor 3, or RUNX3, is a gene that plays a key role in a number of cancers, including gastric neoplasms. With positive expression strongly correlated with better differentiation and a better prognosis, RUNX3 exhibits promise as a prognostic indicator in gastric carcinoma. In human gastric carcinomas, RUNX3 expression tends to gradually decline. RUNX3 transgenic models have been used in studies to show that it can change gastric mucosa into intestinal metaplastic mucosa, with some cases progressing to gastric cancer. RUNX3 is often expressed in gastric cancers and intestinal metaplasia (IM), indicating that it plays a role in the start of gastric carcinogenesis. Developmental genes that promote metastasis in gastric cancer are activated by aberrant upregulation of RUNX3. RUNX3 inhibition in gastric cancer cell lines has been demonstrated in vitro to decrease migration, invasion, and growth independent of anchorage. Furthermore, after splenic inoculation, in vivo studies employing CRISPR-mediated RUNX3-knockout cells show suppression of xenograft growth and liver metastasis. In summary, RUNX3 has a major impact on the onset and spread of gastric neoplasms, suggesting that it could be a promising target for treatment.[21,22]

CD274 - The protein CD274, sometimes referred to as PD-L1 (Programmed Death-Ligand 1), has been found to be a promising target in a number of cancers, including gastric neoplasms. One important target for immune checkpoint blockade treatments is CD274. Inhibitors that target the PD-1/PD-L1 axis have demonstrated exceptional efficacy in inducing tumor regression across various human malignancies. Through its interactions with co-stimulatory and co-inhibitory immune receptors, CD274 plays a critical role in regulating immune responses. In thorough analyses covering a variety of cancer types, CD274 has been found to be a prognostic marker that predicts patient outcomes based on abnormalities in mRNA and protein expression. Microsatellite instability

(MSI), mismatch repair (MMR) status, tumor mutational burden (TMB), drug responsiveness, and the makeup of the tumor immune microenvironment (TIME) are all correlated with aberrant CD274 expression. Interestingly, different levels of immune cell infiltration are linked to different expression levels of CD274. Targeting CD274 is a promising therapeutic approach for cancers that exhibit immune evasion because of its critical role in immune checkpoint dysregulation. To sum up, CD274 has a major role in the development and spread of gastric neoplasms, highlighting its potential as a target for treatment. To completely understand its precise mechanisms and assess its clinical suitability for treating gastric cancers, more research is needed. It is advised to speak with researchers or medical professionals for the most up-to-date and accurate information.[23,24]

MLH1 - MutL Homolog 1, or MLH1, is an important gene in the DNA mismatch repair (MMR) system. It is necessary for finding and fixing mistakes that happen when DNA is copied, such as base-pair mismatches and small nucleotide insertions and deletions. Microsatellite instability (MSI), which is typified by mutations in repetitive DNA sequences, can result from these mistakes if they are not fixed. Because it is linked to particular clinicopathological characteristics, MSI is common in a variety of cancers, including gastric neoplasms, where it serves as a diagnostic and prognostic indicator. The importance of MLH1 mutations as a possible target for therapeutic interventions meant to manage gastric neoplasms is highlighted, especially in Lynch syndrome and sporadic colorectal cancers. To properly understand MLH1's function and take advantage of its therapeutic potential, however, more investigation is necessary. It is recommended to consult recent literature or healthcare professionals for the most up-to-date insights.[25,26]

MET - A tyrosine kinase receptor called MET, or c-Mesenchymal-Epithelial Transition, has been found to be a possible target for a number of cancers, including gastric neoplasms. Tumor invasiveness and a poor prognosis are linked to the activation of the MET/HGF pathway. Increased cell motility, survival, and proliferation can result from activating MET, a receptor for hepatocyte growth factor (HGF). Clinical trials are being conducted to test developed targeted therapies against MET. By blocking the MET/HGF pathway, these treatments seek to stop cancer cells from proliferating and spreading. High levels of MET expression have been linked to a bad prognosis for many types of cancer, such as gastric cancer. As a result, MET may be used as a prognostic indicator. for gastric cancers. In summary, MET may be a target for future treatment approaches because it is crucial to the onset and spread of gastric neoplasms.[1,27]

TSC1 - Tuberous Sclerosis Complex 1, or TSC1, is a gene that is essential for many cancer types. TSC1 has been found to have a significant role in the autosomal dominant disease known as Tuberous Sclerosis Complex (TSC). TSC1 supports development, autophagy, cilia development, cell growth and proliferation, and survival by interacting with several regulatory molecules. It is a crucial part of the pro-survival PI3K/AKT/MTOR signaling cascade. Recent research has shown that TSC1 suppresses tumors in a number of human malignancies, including those of the pancreas, liver, lungs, bladder, breast, and ovaries[2]. TSC1 controls the immune system, migration, invasion, metabolism, and proliferation of cancer cells by identifying inputs from several signaling pathways, including the MTOR/Mdm2/p53 axis, TNF-α/IKK-β, TGF-β-Smad2/3, AKT/Foxo/Bim, and Wnt/β-catenin/Notch. These findings suggest that TSC1 might be a promising therapeutic target for several cancer types.[4,28]

MUC2 - MUC2, a type of stomach gland mucin, has emerged as a prospective target in cancer research due to its significant roles in several cancer types. Being a carrier of the sialyl-Tn antigen in intestinal metaplasia (IM) and the majority of gastric cancer cases demonstrates its significance in tumor growth. Furthermore, MUC2 has been linked to the development of lung invasive mucinous adenocarcinoma.[29,30]

GKN2 - Gastric neoplasms have been found to potentially target GKN2 (Gastrokine 2). It has been discovered that when gastric cancer cells are exposed to oxidative stressors like hydrogen peroxide, their expression of GKN2 rises. GKN2 overexpression encourages mitochondrial dysfunction brought on by reactive oxygen species, which raises cell death. GKN2 triggers apoptosis by activating the JNK signaling pathway and inhibiting the NF-κB signaling pathway. The direct interaction between GKN2 and Hsc70 facilitates this process. As a result, GKN2 may be a useful target for treatment of gastric neoplasms. To completely comprehend its function and potential in the treatment of gastric cancer, more research is necessary. For more specific information, please speak with a healthcare provider.[1,31]

EGFR - The epidermal growth factor receptor (EGFR) is a prospective target for certain malignancies, particularly stomach neoplasms, according to research done after 2022. Important discoveries demonstrate the promise of EGFR-targeted treatments. Due to unselected patient populations, EGFR inhibitors such as cetuximab, panitumumab, and gefitinib did not show any appreciable clinical benefits in esophagogastric cancer despite extensive phase III trials. Finding biomarkers for patients with EGFR-amplified tumors and comprehending the function of EGFR in gastric cancer metastasis in conjunction with MMP7 are the main goals of current research. Furthermore, a promising therapeutic strategy is the combination of monoclonal antibodies and EGFR tyrosine kinase inhibitors. All things considered, EGFR's role in gastric cancer highlights its potential as a target for treatment. [32,33]

BCL2 - A crucial component of the BCL-2 protein family, BCL2 plays a crucial role in controlling apoptotic cell death. Inhibited apoptosis is a common characteristic of many cancers and is frequently caused by dysregulation, whether it be through an overexpression of pro-survival BCL-2 proteins or a decrease in pro-apoptotic counterparts. Although there are few specific research on BCL2's role in stomach tumors published after 2022, the importance of BCL-2 family proteins in apoptosis regulation makes them interesting candidates for the development of cancer therapeutics. The development of BH3-mimetic medications, which block pro-survival BCL-2 proteins, is being investigated as a potential new cancer treatment. These inhibitors target BCL2 and related proteins in an effort to cause cancer cells to undergo apoptosis. The effectiveness of BCL2 targeting for the treatment of gastric cancer, however, varies depending on the tumor's characteristics and the patient's health, highlighting the necessity of individualized medical consultation.[34,35]

TFF2 - In a variety of malignancies, Trefoil Factor 2 (TFF2), a member of the TFF family, acts as a tumor suppressor gene. According to recent studies, DNA methylation, an epigenetic alteration that silences tumor suppressor genes and contributes significantly to the development of gastric cancer, controls the expression of TFF1 and TFF2. Certain CpG island sites have been identified by integrative analyses as being associated with the downregulation of TFF1 and TFF2, suggesting that these sites may be able to inhibit the development of gastric cancer. Thus, TFF1 and TFF2, especially TFF2, may be useful biomarkers for DNA methylation in gastric cancer, indicating that they may be targets for treatment. The significance of individualized medical advice is highlighted by the fact that the role and effectiveness of targeting TFF2 for the treatment of gastric cancer depend on variables like tumor characteristics and patient health.[36]

PSCA - A cell surface protein called prostate stem cell antigen (PSCA) has a variety of uses in various tissues. It was first identified in the prostate, where it can either promote tumor formation or prevent cell division, depending on the situation. In certain tissues, genetic variations of PSCA are linked to an increased risk of developing cancer. PSCA is becoming a prominent target for gastric neoplasms; in vitro research shows that the protein of PSCA is present in gastric neuroendocrine cells, while PSCA mRNA is found in normal gastric mucosa, particularly in the isthmus of gastric glands, but is absent in gastric tumor tissue. Recent studies have demonstrated the efficacy of anti-PSCA CAR-T cells in the fight against gastric cancer, indicating their potential as a treatment option. Personalized medical consultation is crucial because the effectiveness of targeting PSCA in gastric neoplasms depends on variables like tumor characteristics and patient health. [37,38]

TFF1 - A member of the TFF family, Trefoil Factor 1 (TFF1), inhibits tumors in a range of malignancies. Recent studies on gastric cancer have shown that DNA methylation, an epigenetic mechanism that can silence tumor suppressor genes and is known to play a critical role in the progression of gastric cancer, regulates the expression of TFF1 and TFF2. Integrative analyses have revealed the potential of TFF1 and TFF2 to prevent the development of gastric cancer by identifying particular CpG island sites linked to their downregulation. As a result, TFF1 and TFF2 may be a biomarker for DNA methylation in gastric cancer and a potential target for treatment. Personalized medical advice is necessary because the role and efficacy of targeting TFF1 in treating gastric cancer depend on patient health and tumor-specific characteristics. Even though the information in this summary is current, more research is probably needed to fully understand TFF1's involvement in gastric cancer.[36,39]

ARID1A - In gastrointestinal tumors, the AT-rich interaction domain 1 (ARID1A) gene is essential. The protein BAF250a (also known as SMARCF1), which it encodes, is a crucial component of the SWI/SNF chromatin remodeling complex, which controls DNA accessibility by modifying chromatin structure to regulate gene expression. Numerous gastrointestinal cancers, such as colorectal, gastric, and pancreatic cancers, have been found to have mutations in ARID1A. These mutations disrupt the SWI/SNF complex and cause aberrant gene expression, which may be the primary cause of cancer development. ARID1A mutations in gastric cancer are increasingly being identified as important predictive and prognostic factors that impact how well patients respond to therapies such as immunotherapy, PARP inhibitors, mTOR inhibitors, and EZH2 inhibitors. These mutations are becoming useful biomarkers for selective targeted therapies and immune checkpoint blockade therapy. The development of precision therapies is supported by current research that highlights the clinical significance, predictive power, mechanisms, and therapeutic approaches associated with ARID1A mutations in gastric cancer. Even though this summary is based on current research, more investigation is expected to improve our comprehension of ARID1A's function in gastric cancer. For treatment considerations, individualized medical consultation is recommended.[40,41]

S100A8 - A calcium-binding protein belonging to the S100 family, S100A8 is essential for both immune responses and a number of tumor pathologies. This protein, which is mostly made by neutrophils, participates in a variety of signal transduction pathways that are vital for pathogen defense, microtubule organization, and the intricate processes of cancer development, metastasis, drug resistance, and prognosis. S100A8 is a promising candidate for immunotherapy in diffuse large B-cell lymphoma (DLBCL), as it has been demonstrated to induce apoptosis and inhibit tumor growth. Elevated S100A8 levels are associated with poor outcomes and immune infiltration in DLBCL. Although these studies demonstrate the therapeutic potential of S100A8 in a number of cancers, more research is required to determine its function in gastric neoplasms.[42,43]

DPYD - The gene DPYD, or dihydropyrimidine dehydrogenase, is essential for the metabolism of some chemotherapy medications used to treat different kinds of cancer. When it comes to breaking down pyrimidine bases like uracil and thymine, DPYD is the first and rate-limiting enzyme. Additionally, it is essential for the metabolism of fluoropyrimidine medications, including capecitabine and 5-fluorouracil (5-FU), which are frequently used to treat cancer, including gastric cancer. Decreased enzyme activity due to genetic variations in DPYD can result in an accumulation of fluoropyrimidines and increase the risk of severe, potentially fatal toxicity. Consequently, DPYD genotyping has become a useful predictive biomarker for identifying patients who are at risk of fluoropyrimidine therapy side effects. Finding DPYD variations in patients allows for customized fluoropyrimidine dosage modifications, enabling individualized cancer treatment that seeks to improve patient safety and treatment results. In conclusion, even though DPYD may not be a direct therapeutic target for gastric neoplasms, its critical function in drug metabolism highlights its indirect importance in the treatment of cancer. To completely understand DPYD's potential as a therapeutic target in gastric cancer, more research is necessary.[4]

TYMS - Thymidine monophosphate (dTMP), which is essential for DNA replication and repair, is synthesized by the gene TYMS, or thymidylate synthase. TYMS is the target of 5-fluorouracil (5-FU), a chemotherapeutic medication commonly used to treat a range of malignancies, including gastric cancer.

5-FU kills cells by inhibiting TYMS, which halts DNA synthesis. A poor prognosis for various cancer types, including gastric cancer, has been associated with elevated levels of TYMS expression. As a result, TYMS may be used as a prognostic indicator for stomach cancers. The efficacy and adverse effects of chemotherapy regimens based on 5-FU can be impacted by variations in the TYMS gene. As a result, TYMS genotyping has been suggested as a predictive biomarker for customized 5-FU treatments. In conclusion, TYMS plays a crucial role in drug metabolism, which highlights its indirect relevance in cancer management even though it may not be a direct therapeutic target for treating gastric neoplasms. To completely understand TYMS's potential as a therapeutic target, particularly in gastric cancer, more research is required.[1]

CCND1 - Cyclin D1, or CCND1, is a crucial protein that controls the cell cycle and has become a possible target for a number of cancers. CCND1 plays a crucial role in coordinating the cell cycle's transition from the G1 phase to the S phase. Unchecked cell proliferation, a defining feature of cancer, can result from dysregulation of this process. In several cancer types, including gastric cancer, higher levels of CCND1 expression have been linked to a worse prognosis. This implies that CCND1 may be used as a prognostic indicator for stomach cancers. Targeted treatments that block CCND1 are being investigated in clinical trials in an effort to stop the cell cycle and trigger apoptosis. These initiatives highlight CCND1 as a potential target for upcoming gastric cancer treatment approaches. To completely understand CCND1's function and therapeutic potential in gastric neoplasms, however, more investigation is necessary. It is best to speak with researchers or medical professionals for the most up-to-date and accurate information.[1,44]

CTNNB1 - A key player in the Wnt signaling pathway, CTNNB1 (also called β-catenin) has been found to be a possible target for a number of cancers, including gastric neoplasms. The Wnt signaling pathway, which controls cell survival, differentiation, and proliferation, depends on CTNNB1. This system is dysregulated, and in particular, inappropriate activation of Wnt/CTNNB1 signaling leads to unregulated cell proliferation, a characteristic of cancer. Clinical trials are investigating therapies that block CTNNB1 in an attempt to disrupt Wnt signaling and induce cell death. These targeted approaches show that CTNNB1 is a promising target for future cancer treatment strategies, including those for gastric neoplasms. Elevated CTNNB1 expression levels have been linked to poorer outcomes in various cancers, including gastric cancer, suggesting its potential as a prognostic marker. In conclusion, despite the fact that CTNNB1 is not currently directly targeted in the treatment of gastric cancer, its critical function in the Wnt signaling pathway makes it a viable option for upcoming therapeutic interventions.[45]

CDH2 - CDH2, also known as N-cadherin, is a protein crucial for cell-cell adhesion, though literature specifically addressing its role in gastric neoplasms is limited. However, research has implicated CDH2 in other cancer types, where therapeutic strategies targeting its adhesive interactions between cells are under investigation. In contrast, much attention in gastric cancer has focused on CDH1, which is related to CDH2. CDH1 mutations are linked to Hereditary Diffuse Gastric Cancer (HDGC), a condition associated with a high risk of early-onset diffuse gastric cancer (DGC), often requiring prophylactic total gastrectomy due to significant associated morbidity. This underscores the need for alternative treatment approaches. While the direct implications of CDH2 in gastric cancer remain to be fully elucidated, its association with cadherins in cancer underscores its potential significance as a therapeutic target.[46,47]

MSH2 - CTNNB1 (also known as β-catenin), a key protein in the Wnt signaling system, has garnered attention as a potential therapeutic target for a variety of cancers, including gastric neoplasms. Through the Wnt signaling pathway, CTNNB1 is essential for controlling cell survival, differentiation, and proliferation. Uncontrolled cell growth, a defining feature of cancer, is thought to be promoted by dysregulation of this pathway, specifically aberrant activation of Wnt/CTNNB1 signaling. Targeted treatments that block CTNNB1 in order to interfere with Wnt signaling and trigger apoptosis are being studied in clinical trials. These treatment modalities demonstrate that CTNNB1 is a viable option for upcoming anticancer tactics, such as those aimed at gastric neoplasms. CTNNB1 expression has been linked to worse outcomes in a number of cancers, indicating that it may be used as a prognostic indicator for gastric neoplasms as well. In conclusion, even though CTNNB1 isn't a direct target for treating gastric neoplasms at the moment, its critical function in the Wnt signaling pathway makes it a viable target for upcoming therapeutic approaches. To completely define CTNNB1's therapeutic potential, particularly in gastric cancer, more research is needed.[45]

TYMP - Treatment for gastric neoplasms and other cancers may target thymidine phosphorylase (TYMP). Research on colorectal cancer offers some insights, despite the lack of specific studies on gastric neoplasms. TYMP is essential for causing systemic T-cell exhaustion, which can reduce the effectiveness of immunotherapies. Poor clinical outcomes are frequently linked to T-cell exhaustion, a state of T-cell dysfunction that occurs during many chronic infections and cancers. Immunological cell death (ICD), a type of cell death that triggers an immune response against dead cell antigens, has been demonstrated to be induced by targeting TYMP with medications such as tipiracil hydrochloride (TPI)[5]. Tumors may become immunologically "hot" as a result of this process, increasing their propensity to react to immunotherapies. Furthermore, this combined therapeutic method can eradicate tumor-associated macrophages and transform intratumoral regulatory T cells (Tregs) into Th1 effector cells. This results in increased activation and infiltration of cytotoxic T lymphocytes. Furthermore, this impact is connected to the prevention of T-cell fatigue through the reduction of PD-L1 expression in malignancies. In conclusion, TYMP is a promising target for the treatment of gastric neoplasms, which could improve patient outcomes and increase the efficacy of immunotherapies. To completely comprehend TYMP's function in gastric neoplasms and to create efficient TYMP-targeted treatments, more investigation is necessary.[48,49]

VEGFA - One important regulatory gene and angiogenesis marker molecule is vascular endothelial growth factor A (VEGF-A). The development and spread of malignant neoplasms, including gastric neoplasms, are promoted by the upregulation of VEGF-A, which aids in the process of tumor vascularization. In order to stimulate angiogenesis, VEGF-A attaches itself to vascular endothelial growth factor receptors (VEGFR1 or VEGFR2) and triggers subsequent signals. Because it gives tumors the oxygen and nutrients they need, this process is essential. It has been discovered that some genes control the expression of VEGF-A in the context of gastric neoplasms. For example, by controlling VEGFA expression, the cyclic AMP responsive element-binding protein 3-like 4 (CREB3L4) has been demonstrated to stimulate angiogenesis and tumor growth in gastric cancer. The growth and migration of stomach cancer cells brought on by VEGF-A are stopped when CREB3L4 is downregulated. Additionally, VEGF-A has immune-regulating qualities that either directly or indirectly inhibit immune cells' ability to fight tumors. Thus, VEGF-A targeting may improve patient outcomes and increase the efficacy of immunotherapies. The development of VEGF-A-targeted therapy, either by itself or in sensible combinations, has completely changed how many types of cancer are treated. To completely comprehend VEGF-A's function in gastric neoplasms and to create efficient VEGF-A-targeted treatments, more investigation is necessary. [50,51]

CD44 - According to recent research, CD44 is essential for stomach neoplasms. Interestingly, gastric cancer has a substantial overexpression of CD44, indicating that it functions as a tumor stem cell marker in this context. Its potential as a prognostic biomarker for determining the prognosis of gastric cancer is demonstrated by its independent association with immune invasion. Furthermore, CD44 exhibits associations with immune infiltrates, suggesting that it affects survival results. Furthermore, changes in CD44 expression have a direct impact on the behavior of gastric cancer cells, possibly preventing invasion, metastasis, and proliferation. Additionally, changes in CD44 glycosylation can affect how it interacts with ligands and the signaling pathways that follow, which can have a significant impact on the tumor microenvironment and cell fate. Although little is known about the complexities of CD44 glycosylation, more research is necessary to fully understand its crucial role in gastric cancer. All things considered,

CD44 may be a target for stomach cancer diagnosis and treatment; however, further investigation into its glycosylation patterns is required to completely clarify its therapeutic potential.[51,52]

CHGA - Recently, chromogranin A (CGA) has been identified as a key contributor to gastric neoplasms. When combined with EGFR inhibition to improve treatment sensitivity, key findings highlight CGA's function as a prognostic biomarker in gastric cancer, predicting chemoresistance and offering possible therapeutic avenues. According to studies, CGA is widely released from chemoresistant cell lines and is higher in the plasma of chemotherapy patients, especially those who have poor treatment responses and survival rates. Notably, the glycosylation state of CGA determines how it interacts with EGFR, impacting downstream processes like GATA2 activation and, in turn, strengthening chemoresistance mechanisms in gastric cancer via a positive feedback loop. These findings suggest that combining therapeutic approaches that target both CGA and EGFR may help overcome chemoresistance in gastric cancer; however, more investigation is necessary to fully understand CGA's specific function in this disease setting.[1]

RNF180 - Recent studies have shown that Ring Finger Protein 180 (RNF180) plays a major role in gastric neoplasms. The start and progression of gastric cancer are closely linked to methylation-induced decreased expression of RNF180, a tumor suppressor gene necessary for cell growth, proliferation, and apoptosis. RNF180 is thus positioned as a possible gastric cancer diagnostic biomarker. RNF180's diagnostic potential has been highlighted by the discovery that methylation of the protein occurs much more frequently in plasma samples from patients with gastric cancer than in controls. When RNF180 was evaluated as a diagnostic marker, it outperformed conventional biomarkers like CEA, CA199, and CA724, with sensitivity of 0.54, specificity of 0.80, positive likelihood ratio of 2.73, and negative likelihood ratio of 0.58. Moreover, in gastric cancer tissues, there is an inverse relationship between RNF180 expression levels and lymph node metastases. RNF180's function in preventing the potential for metastasis was highlighted in experimental models where its re-expression inhibited tumor growth, lymphangiogenesis, and the activation of malignant molecular signaling. These results imply that RNF180 has potential as an accurate target for gastric cancer diagnosis and treatment.[53,54]

GACAT3 - Recent studies have shown that Gastric Cancer Associated Transcript 3 (GACAT3) plays a major role in gastric neoplasms. GACAT3, a long non-coding RNA (lncRNA) with more than 200 nucleotides, is mainly involved in the regulation of gene expression and epigenetics. GACAT3, an oncogenic lncRNA, has been found to be dysregulated in a variety of tumors, impacting clinical features and fostering a number of oncogenic processes. GACAT3 is a promising biomarker for diagnosis, prognosis, and possible targeted therapies because of its involvement in various cancer types. GACAT3's therapeutic importance in the treatment of gastric cancer is further supported by the fact that its role as an inflammatory response gene has been connected to promoting the proliferation of gastric cancer cells. Remarkably, distinct expression patterns of GACAT3 have been seen in various gastric cancer cell lines, underscoring its complex regulatory function in the progression of the illness. Although GACAT3 exhibits promise as a precise target for the treatment of gastric cancer, more investigation is necessary to completely clarify its comprehensive role in disease mechanisms.[55]

CDKN1A - Recent studies have shown that Cyclin-Dependent Kinase Inhibitor 1A (CDKN1A), commonly referred to as p21, plays a significant role in the context of gastric neoplasms. Important discoveries point to its possible role in gastrointestinal cancers, as evidenced by research that links CDKN1A polymorphisms to an increased risk of esophageal cancer in relation to age. A pan-cancer analysis also showed that changes in CDKN2A, a gene linked to CDKN1A, were linked to the inhibition of multiple immune-related pathways in gastric cancer, such as interferon alpha and gamma responses, indicating a function for CDKN1A in regulating immune responses in gastric cancer. Additionally, this analysis showed that changes in CDKN2A were associated with worse outcomes in a variety of cancers, suggesting that CDKN1A may be a prognostic marker for gastric cancer. Despite these discoveries, more investigation is necessary to completely clarify the therapeutic potential of CDKN1A and its precise function in gastric neoplasms.[48,56]

COMMD6 - Numerous malignancies have been linked to Copper Metabolism Domain Containing 6 (COMMD6). In various cancer types, including as cholangiocarcinoma, adrenocortical carcinoma, and head and neck squamous cell carcinoma, elevated COMMD6 expression has been associated with decreased overall survival (OS) and disease-free survival (DFS). Despite these correlations, there is a dearth of thorough studies explicitly examining COMMD6's function in gastric neoplasms. Future research may identify COMMD6 as a potential target as our knowledge of the molecular mechanisms underlying gastric cancer develops, requiring more examination of its implications in gastric cancer pathology.[1,57]

FGFR2 - With novel strategies like proteolysis-targeting chimeras (PROTACs) being investigated for its therapeutic targeting, FGFR2 has become a crucial target in gastric cancer. For FGFR2b-selected gastric or gastro-oesophageal junction adenocarcinoma, bemarituzumab, a novel afucosylated humanized IgG1 anti-FGFR2b monoclonal antibody, has been studied in conjunction with a modified mFOLFOX6 regimen. Early detection of gastric cancer and intestinal metaplasia may be possible due to the non-invasive biomarker FGFR2 methylation, which can be found in blood leukocytes. Interestingly, in gastric adenocarcinoma, FGFR2 expression or amplification plays a crucial role in patient selection for targeted therapies; for example, bemarituzumab is more effective in patients whose FGFR2 expression is identified by immunohistochemistry. Research is necessary to completely understand FGFR2's complex role in gastric cancer pathology, even though its role in immune infiltration and gastric cancer prognosis highlights its potential as a precise therapeutic target.[58–60]

VIM - Numerous malignancies have been linked to vitexin (VIM). In various cancer types, increased VIM expression has been associated with better tumor development and metastasis, indicating that it may be a valuable target for cancer therapy. To clarify its precise function in gastric cancer, however, more thorough investigation is required. Although preliminary data points to VIM's involvement in gastric neoplasms, more research is necessary to completely evaluate its potential as a therapeutic target.[1]

CEACAM5 - According to recent research, gastric neoplasms are significantly influenced by Carcinoembryonic Antigen-Related Cell Adhesion Molecule 5 (CEACAM5). Normal tissues like the colon, esophagus, and head and neck do not express the cell surface glycoprotein CEACAM5, but it is frequently overexpressed in a variety of tumor types, such as lung, breast, and gastrointestinal cancers. Recent research has shown that CEACAM5 is a promising therapeutic target, especially in gastric cancer, where Liu Nian's team's studies clarified how it promotes the migration, proliferation, and epithelial-mesenchymal transition (EMT) of cancer cells. In the context of incomplete intestinal metaplasia, our study also highlights CEACAM5 and TROP2 as possible molecular markers for identifying people who are more likely to develop gastric cancer. Additionally, CEACAM5-positive non-squamous non-small cell lung cancer may benefit from increased treatment efficacy when CEACAM5 is targeted with the antibody-drug conjugate tusamitamab ravtansine (SAR408701). Although CEACAM5 shows promise as an accurate target for gastric cancer diagnosis and treatment.[61,62]

MYC - According to recent research, gastric neoplasms are significantly influenced by MYC, a protein that is frequently dysregulated in cancer. MYC is a family of transcription factors that includes c-MYC, L-MYC, and N-MYC. These transcription factors are master regulators of important cellular functions like metabolism, differentiation, proliferation, and the cell cycle. Despite past difficulties in creating safe and efficient inhibitors, MYC is a desirable therapeutic target because overactivity of these proteins can promote carcinogenesis and maintain tumor growth[2]. Effective methods for addressing MYC have been shown in recent research. For example, studies have shown that the conserved eukaryotic protein ZC3H15, through its modulation of the FBXW7/c-Myc pathway, promotes the progression of gastric cancer. In patients with gastric cancer, high ZC3H15 expression is associated with a lower chance of survival because it increases c-Myc expression, which promotes cell invasion, migration, and proliferation. Additionally, research on how MYC interacts with microRNAs in gastric cancer seeks to find possible therapeutic targets and biomarkers. Even though MYC appears to be a promising target for accurate diagnosis and treatment of gastric cancer, more study is necessary to completely understand its complex role in the development and management of the disease.[63,64]

MMP7 - Recent studies have shown that matrix metallopeptidase 7 (MMP7) plays a major role in gastric neoplasms. Important discoveries highlight MMP7 and EGFR as key players in gastric cancer metastases, with significant overexpression seen in gastric adenocarcinoma strongly associated with poor patient outcomes. Research indicates that MMP7 overexpression is a common feature of many human cancer types, including stomach cancer, and that it is essential for encouraging the invasion, migration, and multiplication of cancer cells. In order to stop the growth and spread of gastric adenocarcinoma, MMP7 targeting appears to be a promising treatment strategy. Although MMP7's role in the prognosis and metastasis of gastric cancer indicates its potential as an accurate target for diagnosis and treatment.[65,66]

ZNF135 - Although there are few direct connections between zinc finger protein 135 (ZNF135) and gastric neoplasms, ZNF135 has been identified as a significant factor in a number of cancers. ZNF135 may be a promising target for cancer treatment because it has been associated with better tumor growth and metastasis in other cancer types. Further research is required to elucidate the precise role of ZNF135 in gastric cancer. ZNF135 may be involved in gastric neoplasms, according to preliminary data.[48]

STAT3 - Recent studies have shown that Signal Transducer and Activator of Transcription 3 (STAT3) plays a significant role in gastric neoplasms. One transcription factor that has been identified as a potential target for the development of anticancer medications in gastric cancer is STAT3. But as of yet, no STAT3 inhibitor has received FDA approval. According to recent research, STAT3 can be successfully targeted. For example, proteolysis-targeting chimeras (PROTACs) have been used to design and synthesize a class of STAT3 degraders. Among these, SDL-1 degrades the STAT3 protein in vitro, has strong anti-gastric cancer cell proliferation properties, prevents MKN1 cell invasion and metastasis, triggers MKN1 cell apoptosis, and simultaneously stops the cell cycle. One possible treatment target for gastric cancer is CD163, a novel STAT3 target gene. A poor prognosis and tumor invasion are linked to high CD163 expression. In vivo tumor growth may be inhibited by knocking down CD163 in cancer cells. The formation of distal liver metastases is encouraged by elevated STAT3 signaling in the tumor microenvironment. To sum up, STAT3 is important in gastric cancer, and because of its strong correlation with prognosis and immune infiltration, it may be a precise target for gastric cancer diagnosis and treatment.[67,68]

RPRM - Numerous tumors have been discovered to be significantly influenced by Reprimo (RPRM). As a tumor-suppressor gene in other cancer types, RPRM has been linked to the development of several malignant tumors, including gastric cancer and pituitary tumors. As a result, it may be a biomarker for early cancer detection.

Although some evidence points to RPRM's potential involvement in gastric neoplasms, further investigation is required to completely comprehend its potential as a therapeutic target.[48,69]

PDCD1 - Research on gastric cancer has focused on PD-1, or programmed cell death protein 1, as an immune checkpoint that can inhibit T-cell activity by binding with its ligand, PD-L1. The promise of immune checkpoint blockade therapies in the fight against cancer has been highlighted by the significant effectiveness of monoclonal antibodies that target PD-1 in the treatment of tumors. However, resistance problems still exist, and current applications are mostly limited to treating advanced tumors following initial treatment failures. In an effort to improve therapeutic outcomes based on synergistic effects seen in preclinical models, recent research has investigated combining anti-PD-1 therapies with now-available treatments for HER2-positive gastric cancers, such as trastuzumab and chemotherapy. Furthermore, studies on PD-1's post-translational modifications, such as glycosylation and phosphorylation, have demonstrated their crucial role in controlling PD-1's activities, opening up new ways to strengthen anti-tumor immune responses and combat PD-1-mediated immune suppression.[70,71]

AFP - One oncofetal glycoprotein that has been investigated as a possible target for gastric cancers is alpha-fetoprotein (AFP). A rare form of gastric cancer with a high degree of malignancy, a high frequency of metastases, and a poor prognosis is called AFP-positive gastric cancer (AFPGC). By suppressing natural killer (NK) cells and adversely controlling dendritic cell activity, AFP may have an impact on immunity. According to a 2024 study, baseline serum AFP levels may be able to predict how well immune checkpoint inhibitors (ICIs) will work for patients with advanced gastric cancer (AGC). Higher baseline AFP levels were linked to a decline in the efficacy of ICIs, according to the study's retrospective analysis of 158 AGC patients who received ICI treatment.[72]

MMP9 - One secretory endopeptidase that has been found to be an essential mediator of processes closely linked to carcinogenesis is matrix metalloproteinase 9 (MMP9). By controlling the migration, epithelial-to-mesenchymal transition, and survival of cancer cells, as well as by inducing the immune response, angiogenesis, and the formation of the tumor microenvironment, MMP9's proteolytic activity plays a crucial role in tumorigenesis. It appears that MMP9 is a desirable target for anticancer treatments. However, because MMP9's specific, frequently contradictory role is determined by a very complex mechanism of regulation of expression, synthesis, and activation, it is very difficult to develop safe and effective MMP9 inhibitors as anticancer drugs. Furthermore, MMP9 shares a high degree of homology with other MMP family members. Blocking antibodies that specifically deactivate MMP9 have recently generated excitement and are presently undergoing clinical trials. We still need to learn more about the mechanisms by which MMP9 contributes to the carcinogenesis process, though.[73]

CASP3 - A particular protease called CASP3, or caspase-3, cleaves substrates like acetyl-DEVD-7-amino-4-methylcoumarin and poly-ADP ribose polymerase. Apoptosis is characterized by DNA fragmentation, which is caused by this enzymatic activity. According to recent research, CASP3 promotes apoptosis, which is essential for tumor suppression. A rigorous pan-cancer investigation using the Genotype-Tissue Expression and Cancer Genome Atlas databases revealed a substantial correlation between CASP3 expression and the prognosis of most malignancies. Furthermore, in almost all tumor types, there was a correlation between CASP3 expression and the tumor microenvironment. B cell activation, antigen presentation, immunological responses, chemokine receptors, and inflammatory function are all likely involved in CASP3 action in addition to apoptosis. CASP3 was implicated in tumor invasion and metastasis in a study on colon cancer, and its deletion frequently results in increased sensitivity to radiation and chemotherapy, indicating that cleaved CASP3 could be a novel target for cancer treatment.[74]

## 1.2. Gastric Neoplasm treatment:

*Targeted Therapy:*

Ramucirumab - Based on a number of important factors, ramucirumab is a good treatment option for gastric neoplasms. It is frequently used as a second-line targeted therapy for advanced gastric or gastro-esophageal junction (GEJ) adenocarcinoma after fluoropyrimidine or platinum-based chemotherapy.

Ramucirumab, a fully human monoclonal antibody (IgG1), is categorized as a biotech medication and is designed to fight solid tumors. It works by directly blocking the trans-membrane tyrosine kinase receptor VEGFR2, which is present on endothelial cells. By binding to VEGFR2 and blocking the receptor's interaction with its ligands (VEGF-A, VEGF-C, and VEGF-D), ramucirumab reduces VEGF-induced signaling, which is crucial for angiogenesis and the progression of cancer. Its effectiveness in improving overall survival and investigating combination therapies in advanced gastric cancer treatment is demonstrated by recent studies like the RAINBOW and RAP trials. To guarantee the best possible treatment choice, the choice to use Ramucirumab should be customized based on the unique circumstances of each patient and discussed with medical professionals.[75,76]

Trastuzumab - When treating gastric neoplasms, especially those with HER2-positive status, trastuzumab is essential. It is mostly used as a targeted treatment for HER2-positive locally advanced, incurable, or metastatic gastric or gastroesophageal junction (GEJ) adenocarcinomas. It is frequently used in conjunction with pembrolizumab and chemotherapy. One example of a biotech drug is trastuzumab, a humanized monoclonal antibody derived from mice. It slows cell division and produces antibody-dependent cellular cytotoxicity by binding to the extracellular domain of the HER2 receptor. Furthermore, trastuzumab has been demonstrated to promote HER2 internalization and degradation by enhancing the activity of the tyrosine kinase–ubiquitin ligase c-Cbl pathway. Despite being the first-line treatment for HER2-positive malignancies, including gastric cancer, trastuzumab's efficacy may be affected by problems such as resistance mechanisms involving many proteins and signaling pathways. In order to maximize trastuzumab's therapeutic benefits in the treatment of gastric cancer, ongoing research attempts to resolve these complexities.[1,77]

Rivoceranib - Riveranib is also known as apatinib, a small-molecule tyrosine kinase inhibitor (TKI) that has been investigated for the treatment of gastric neoplasms. Riviveranib specifically blocks vascular endothelial growth factor receptor-2 (VEGFR-2), a critical mechanism in tumor angiogenesis where tumors form new blood vessels for sustenance. By inhibiting VEGFR-2, Rivoceranib is believed to stop tumor development and disease progression by restricting VEGF-mediated activities such endothelial cell migration and proliferation, which stops new blood vessels from forming in tumor tissues. Rivoceranib uses a targeted therapy strategy, specifically targeting and blocking pathways that are essential for the growth and progression of tumors, particularly the VEGFR-2 pathway. Rivoceranib's effectiveness in treating gastric cancer has been the subject of recent studies with promising results. Research on gastric cancer patients showed encouraging outcomes, including a phase 2 trial that was carried out in South Korea and the United States. While overall survival benefits were not statistically significant, the CARES-310 trial showed nearly doubled progression-free survival. Ongoing research, however, emphasizes the intricacy of treatment results and the necessity of individualized medical guidance and treatment alternatives in clinical practice. Consulting medical experts is crucial for individualized information and treatment recommendations.[78]

Oxonic Acid Drug - A tiny molecule called oxonic acid, sometimes referred to as oteracil, is used in conjunction with cancer treatments to lessen gastrointestinal toxicity. It is a major ingredient in "Teysuno," a drug that combines oxonic acid, gimeracil, and tegafur. The main active component of Teysuno, tegafur, functions as a pro-drug of fluorouracil (5-FU), a strong anti-metabolite that targets rapidly dividing cancer cells by integrating into their DNA and RNA strands and preventing their replication. Oxonic acid acts as a targeted therapy by inhibiting orotate phosphoribosyltransferase (OPRT), an enzyme necessary for the production of 5-FU. By decreasing 5-FU's activity in healthy gastrointestinal tissue, this modulation lessens the harmful effects that are linked to it. Numerous treatments, such as chemotherapy, radiotherapy, surgery, immunotherapy, and targeted therapy, have shown effectiveness against gastric adenocarcinoma in relation to its use in gastric neoplasms. Thus, depending on the specific patient and tumor characteristics, oxonic acid may be useful in the treatment of gastric cancers when used in a targeted therapy approach.[1,79]

### *Chemotherapy:*

Tegafur-gimeracil-oteracil Potassium - Teracil-tegafur-gimeracil Advanced stomach cancer can be treated with a drug called teysuno, which is another name for potassium. A number of cancers, such as colorectal, breast, pancreatic, biliary tract, non-small-cell lung, and head and neck cancers, are treated with it in combination with cisplatin. Teracil-tegafur-gimeracil Potassium falls under the category of tiny molecules. Three pharmaceutical ingredients are combined to form it: oteracil, gimeracil, and tegafur.

Tegafur, a pro-drug of fluorouracil (5-FU), a cytotoxic anti-metabolite medication that targets rapidly dividing cancer cells, is the primary active ingredient in Teysuno. 5-FU can insert itself into DNA and RNA strands by imitating a class of molecules known as "pyrimidines" that are vital to RNA and DNA. This stops the replication process that is required for cancer to continue growing. By reversibly inhibiting the dehydrogenase enzyme dihydropyrimidine dehydrogenase (DPD), gemeracil prevents fluorouracil from degrading. Higher 5-FU levels and a longer half-life of the drug are the outcomes of this. The main way that oteracil in Teysuno reduces gastrointestinal toxicity is by lowering 5-FU activity in the healthy gastrointestinal mucosa. It works by preventing the production of 5-FU by the enzyme orotate phosphoribosyltransferase (OPRT). Chemotherapy is the treatment approach for tegafur-gimeracil-oteracil potassium. It is applied in addition to antitumor therapy. When combined with cisplatin, it is approved for the treatment of advanced gastric cancer in the European Union.[80,81]

Tegafur - Based on a number of criteria, tegafur is an appropriate medication for gastric neoplasms. Tegafur is a component of a chemotherapy treatment plan. It is frequently administered in conjunction with medications like gimeracil and oteracil that increase its bioavailability and reduce its toxicity. For instance, it is advised to treat advanced stomach cancer in conjunction with cisplatin. Tegafur is a tiny chemical that is used as a medicine. Small compounds can easily cross the cell membrane to enter intracellular areas of action. Tegafur is a prodrug of the antitumor drug 5-fluorouracil (5-FU). Tegafur is mainly metabolically activated in the liver after administration, where it is changed into 5-FU. By blocking thymidylate synthase (TS) during the pyrimidine pathway involved in DNA synthesis, the active 5-FU mediates an anticancer effect. Tegafur works therapeutically by interfering with DNA synthesis, which is essential for the development and spread of cancer cells.[81]

Fluorouracil - Fluorouracil (5-FU), according to recent research, is an effective drug for treating stomach neoplasms. This chemotherapy drug is commonly used in combination treatments, like the FLOT regimen (fluorouracil, leucovorin, oxaliplatin, and docetaxel), which is the gold standard for treating gastric cancer during surgery. Fluorouracil is a pyrimidine analog antimetabolite and a small molecule cytotoxic chemotherapy drug. By imitating RNA and DNA building blocks, it interferes with the synthesis of proteins and DNA, which slows or stops the growth of cancer cells. Thymidine triphosphate, which is essential for DNA synthesis, is depleted as a result of its active metabolite, F-dUMP, inhibiting thymidylate synthase. Individual responses can differ, so seeking advice from medical professionals is necessary to determine the best course of action, even though fluorouracil has shown promise in treating gastric neoplasms.[82,83]

Oxaliplatin - A number of important considerations make oxaliplatin an appropriate treatment for gastric neoplasms. Mostly used in chemotherapy, it is commonly used in conjunction with other drugs like capecitabine in the CAPOX regimen or fluorouracil and leucovorin in the FOLFOX regimen[125]. The platinum-based chemotherapy family, which also contains cisplatin and carboplatin[5], includes the small molecule drug oxaliplatin. By acting as an alkylating agent and incorporating platinum metal into DNA, it hinders DNA replication and produces cytotoxic effects, which in turn slows or stops the growth of cancer cells and encourages cell death. Studies showing increased survival rates when Oxaliplatin is incorporated into treatment protocols for advanced gastric cancer, particularly in combination therapies catered to patient needs, highlight the drug's effectiveness in treating gastric neoplasms. However, the choice to employ oxaliplatin should be discussed with medical experts and should take into account the unique circumstances of each patient as well as the characteristics of the tumor.

Cisplatin - In chemotherapy, cisplatin is commonly used to treat a range of malignancies, including stomach neoplasms. It is typically used in combination with other drugs to treat advanced gastric cancer. Cisplatin is a small molecule medication that is categorized as an antineoplastic agent based on platinum. In order to hinder DNA replication and function, it binds to DNA, creates cross-links, breaks the double helix structure, and covalently binds to DNA bases. This damage is permanent in terms of cell division, which prevents or slows the proliferation of cancer cells and ultimately leads to cell death. The development of resistance mechanisms involving different proteins and signaling pathways can jeopardize the clinical utility of cisplatin, even though it is an effective first-line chemotherapy. In order to maximize the effectiveness of cisplatin in the treatment of gastric cancer, ongoing research aims to understand these mechanisms and investigate ways to reduce resistance.[84,85]

Capecitabine - Because of its unique properties and mode of action, capecitabine is regarded as a useful treatment option for gastric neoplasms. Capecitabine is a chemotherapeutic agent that is classified as a small molecule drug. It is commonly used in combination therapy in order to maximize its effectiveness. It works as a prodrug that the body converts into 5-fluorouracil (5-FU), an antimetabolite that stops DNA, RNA, and proteins from being synthesized. This stops malignant and quickly dividing cells from growing and eventually kills them. When used in conjunction with Oxaliplatin or Cadonilimab, Capecitabine has been shown in recent studies to improve survival outcomes for patients with advanced gastric or gastroesophageal junction adenocarcinoma. Furthermore, studies that contrast Capecitabine plus cisplatin with conventional therapies have revealed similar progression-free survival rates for the treatment of gastric cancer. However, choices about Capecitabine therapy should be made in consultation with medical experts, taking into account specific patient characteristics that could affect the safety and effectiveness of treatment.[1]

FLOT Regimen - Fluorouracil, leucovorin, oxaliplatin, and docetaxel, or FLOT, is a chemotherapy regimen that has demonstrated significant effectiveness as perioperative treatment for patients with locally advanced gastric (GC) and gastroesophageal (AEG) carcinoma. One perioperative chemotherapy strategy is the FLOT regimen. This indicates that it is given both prior to (neoadjuvant) and following (adjuvant) surgery. The FLOT cycles before surgery can help shrink the tumor, and the cycles after surgery can help get rid of any remaining cancer cells. A variety of small molecule medications make up the FLOT regimen. Every element of the FLOT regimen plays a distinct part in the treatment of cancer. An antimetabolite medication called fluorouracil (5-FU) destroys cancer cells by interfering with DNA replication and functioning as fake building blocks in the cell's genes. Leucovorin is administered to enhance the effectiveness of fluorouracil; it is not a chemotherapy medication.

Oxaliplatin is an alkylating agent that stops a cell from dividing into new cells by interfering with the DNA's ability to develop. The antineoplastic medication docetaxel contains chemicals that kill cancer cells and other cells that divide quickly. The FLOT regimen is currently the gold standard for treating fit patients with operable gastric (GC) or gastroesophageal (GEJ) adenocarcinomas, per recent studies. Its 5-year overall survival (OS) is 45%, which is higher than the 23% achieved with surgery alone. Patients treated with FLOT had a statistically significant longer median overall survival of 57.8 months compared to 28.9 months, according to another study. The viability of FLOT in an unselected population that is representative of clinical practice is confirmed by these empirical data.[82,86]

Methylnitronitrosoguanidine - A tiny molecule known as methylnitronitrosoguanidine (MNNG) has been employed in experiments as a mutagen and carcinogen. It works by introducing alkyl groups to guanine's O6 and thymine's O4, which may result in transition mutations from GC to AT. This mechanism of action results from the degradation of its electrophilic compounds, where the mutagen ammonium ion causes an electrophilic effect on DNA base pairs, particularly on its nucleophilic sites. Many treatment approaches, such as surgery and the use of glucagon-like peptide 1 (GLP-1) receptor agonists, are being investigated for gastric neoplasms. The particular kind and stage of the tumor, as well as the patient's general health, determine whether a treatment is appropriate. For individualized medical advice, it is imperative to speak with healthcare professionals.[87,88]

Fluoropyrimidine - One type of chemotherapy used to treat a variety of solid tumors, including gastric cancers, is fluoropyrimidine, which includes 5-fluorouracil (5-FU) and capecitabine. It falls under the category of small molecule drugs. Fluoropyrimidine is commonly used in clinical practice in conjunction with other drugs to treat gastric cancer. For example, the FDA has approved the use of fluoropyrimidine and platinum-based chemotherapy in conjunction with the immunotherapy medication nivolumab for advanced gastric cancer. As a third-line adjuvant treatment for advanced gastric or gastroesophageal junction tumors, recent research has also investigated the combination of immunotherapy and the tyrosine kinase inhibitor apatinib[11]. When used as a prodrug or in combination with a dihydropyrimidine dehydrogenase inhibitor, fluoropyrimidine works by preventing RNA synthesis and function, reducing the activity of thymidylate synthase, and integrating into DNA to cause breaks in DNA strands. These mechanisms demonstrate how effective it is at treating cancer, but medical professionals should always guide treatment decisions that are specific to each patient's needs and tumor characteristics.[1,89–93]

Docetaxel - According to recent studies and its unique properties, docetaxel is regarded as a promising treatment option for gastric neoplasms. Docetaxel is mostly used in chemotherapy, frequently in combination with other medications such as S-1, or as a component of postoperative care. Docetaxel is a member of the taxane family of drugs and is categorized as a small molecule. Docetaxel stops cell division by interfering with normal microtubule function in cells. It enhances microtubule assembly and prevents disassembly by binding to tubulin, an essential component of microtubules. This stabilization process inhibits tumor growth and disrupts mitosis, which ultimately results in cell death. Docetaxel and S-1 work well together, according to recent studies. The international Phase III START trial's long-term follow-up showed that the Docetaxel plus S-1 regimen considerably improved patient outcomes when compared to S-1 alone. Furthermore, research has shown that patients with gastric or gastro-esophageal cancer may benefit greatly from docetaxel-based combinations during the perioperative phase. Furthermore, a randomized phase 3 trial showed that patients with operable gastric cancer who received perioperative chemotherapy based on docetaxel had an improved overall survival rate. For personalized medical advice, it's essential to consult with a healthcare professional. The information provided is based on recent studies and general knowledge about Docetaxel; individual responses to treatment may vary.[94,95]

Leucovorin - Recent research and its mechanism of action support the use of leucovorin in the treatment of gastric neoplasms. Leucovorin is an essential part of chemotherapy, especially when combined with fluorouracil and oxaliplatin in the FOLFOX regimen. Although this treatment works well at first, patients may eventually become resistant to it, which could cause their tumor to grow. Leucovorin is also used to treat colorectal cancer in conjunction with medications such as 5-fluorouracil. Leucovorin is categorized as a small molecule. Folic acid derivative leucovorin supports cellular functions as an active metabolite that doesn't need to be activated by dihydrofolate reductase. It is a vital coenzyme in the synthesis of nucleic acids, which is vital for cellular activity. Leucovorin is used in clinical settings to improve the efficacy of fluorouracil against cancer cells and lessen the harmful effects of methotrexate on healthy cells. Although leucovorin is used to treat gastric neoplasms, its effectiveness varies and resistance may grow over time. For individualized treatment plans, speaking with a healthcare professional is advised.[1,96]

Mitomycin - Based on new research and its mechanism of action, mitomycin is used to treat gastric neoplasms. As a chemotherapeutic agent, mitomycin is frequently used in combination with medications such as doxorubicin and 5-fluorouracil to treat a variety of cancers. Ongoing research explores its potential in cytoreductive surgery for advanced gastrointestinal malignancies. One type of small molecule is mitomycin. By cross-linking the complementary strands of the DNA double helix, mitomycin acts as an alkylating agent that prevents DNA synthesis. Because of this interference, DNA transcription into RNA is hindered, which stops protein synthesis and prevents the cancer cell from proliferating. It also prevents the synthesis of proteins and RNA at higher concentrations. Although mitomycin is used to treat gastric neoplasms, its effectiveness varies and resistance may eventually arise. For individualized treatment options, patients are encouraged to speak with medical professionals.[94,97]

Irinotecan - Recent research and its mode of action support the use of irinotecan in the treatment of gastric neoplasms. Irinotecan is frequently used as a chemotherapeutic drug in combination with leucovorin and 5-fluorouracil to treat a range of malignancies. Research is also investigating its potential in cytoreductive surgery for advanced gastrointestinal malignancies. Recent studies highlight its utilization alongside immune checkpoint inhibitors following the failure of second-line paclitaxel plus ramucirumab treatment in patients with locally advanced unresectable or metastatic advanced gastric cancer. Irinotecan is classified as a small molecule. Irinotecan functions as a topoisomerase inhibitor, specifically blocking the activity of the topoisomerase I enzyme. This activity damages DNA, which leads to cell death. By binding to topoisomerase I, blocking its activity, and interfering with DNA synthesis, irinotecan and its active metabolite SN-38 both have antitumor effects that stop the cell cycle in the S-G2 phase. Although irinotecan is a good treatment for gastric neoplasms, its effectiveness varies from person to person and resistance may grow over time. For individualized treatment plans, patients are encouraged to speak with healthcare professionals.[98,99]

FAMTX Regimen - The FAMTX regimen, comprising 5-fluorouracil, adriamycin (doxorubicin), and methotrexate, stands out as a highly effective combination for combating gastric cancer. FAMTX is a chemotherapy regimen, leveraging a blend of cytotoxic drugs that synergistically target and eradicate cancer cells. These medications are classified as small molecules, capable of penetrating cells and acting on intracellular targets. FAMTX's constituents impede the growth and division of cancer cells. In particular, 5-fluorouracil and methotrexate stop cell proliferation by blocking thymidylate synthase, which is essential for DNA synthesis. Adriamycin works by intercalating into DNA, causing structural disruption and preventing replication. New studies show that altering the FAMTX regimen, such as lowering the dosage of methotrexate, increases its effectiveness in treating advanced gastric cancer. In clinical trials involving patients with unresectable locally advanced or metastatic gastric carcinoma, this adjusted regimen yielded a notable total response rate of 47%. The median response duration exceeded 8 months, with an overall median survival of 10 months. These findings underscore the FAMTX regimen's potential as a viable treatment option for gastric neoplasms. However, a number of variables, such as the patient's health, the stage of the cancer, and the genetic makeup of the tumor, influence the choice of treatment. As a result, seeking advice from medical professionals is crucial in order to customize the best course of action for each patient's unique situation.[100,101]

CAPOX Regimen - Gastric neoplasms are commonly treated with the CAPOX regimen, which consists of capecitabine and oxaliplatin. CAPOX is a type of chemotherapy that consists of cytotoxic medications that work together to kill cancer cells. These drugs fall into the category of small molecules because they can enter cells and target intracellular processes. CAPOX works by causing DNA damage in cells that divide quickly, especially cancer cells. The body converts capecitabine to 5-fluorouracil, which inhibits an enzyme essential for DNA synthesis and stops the division of cancer cells. By creating cross-links with DNA, oxaliplatin prevents its transcription and replication, which eventually results in cell death. The effectiveness of the CAPOX regimen in treating gastric cancer has been highlighted by recent studies. Combining sugemalimab with CAPOX improved outcomes for patients with advanced, untreated gastric or gastroesophageal junction adenocarcinoma, according to studies presented at the ESMO Congress 2023. Additionally, comparative studies have shown CAPOX to be more effective than the S-1 regimen for treating stage IIIB or IIIC gastric cancer. However, a number of variables, including the patient's general health, the stage of the cancer, and the genetic makeup of the tumor, affect treatment choices. In order to choose the best course of treatment, patients and healthcare professionals must work closely together.[102,103]

Hyperthermic Intraperitoneal Chemotherapy - Gastric neoplasms can be treated with hyperthermic intraperitoneal chemotherapy (HIPEC). HIPEC is a type of chemotherapy used to treat advanced abdominal cancers that combines surgery and hyperthermia therapy. The medications used in HIPEC are categorized as small molecules because they can enter cells and target intracellular components. Chemotherapy medications are heated to temperatures between 108 degrees Fahrenheit (41 and 43 degrees Celsius) and briefly circulated within the peritoneal cavity during HIPEC. The drugs' cytotoxic effects are intensified by the higher temperature, which efficiently destroys cancer cells—especially the microscopic ones found in the abdominal cavity. The effectiveness of HIPEC in treating gastric neoplasms has been demonstrated by recent studies. Research has demonstrated the safety and potential of HIPEC in the treatment of gastric cancer with peritoneal metastases, including a single-center Chinese study. Additionally, studies conducted by organizations such as the Mayo Clinic have shown that HIPEC is more effective than traditional chemotherapy in treating stomach cancer, and ongoing research is being conducted to assess its wider applicability. However, a number of variables, including the patient's general health, the stage of the cancer, and the genetic makeup of the tumor, influence treatment choices. Therefore, in order to choose the best course of treatment, patients must have in-depth conversations with healthcare professionals.[104,105]

Doxifluridine - One medication being researched for the treatment of gastric neoplasms is doxifluridine. Recent research emphasizes its type, mechanism of action, treatment approach, and possible efficacy. Doxifluridine has demonstrated promise as a chemotherapy treatment for patients with advanced or recurrent gastric cancer when combined with other medications, such as Paclitaxel. This combination is a good outpatient treatment plan because it is well-tolerated and shows efficacy. Doxifluridine is categorized as a small molecule. Within tumor cells, the fluoropyrimidine derivative doxifluridine transforms into 5-fluorouracil (5-FU). The enzyme pyrimidine nucleoside phosphorylase (PyNPase), which is more active in tumors than in healthy tissues, mediates this conversion. As a result, tumor tissues contain higher concentrations of 5-FU, which is selectively cytotoxic. Although doxifluridine has shown some efficacy, each patient will react differently. Both the type of gastric neoplasm and the unique features of the patient must be taken into account. For individualized medical advice, speaking with a healthcare professional is advised.[106,107]

Epirubicin - One chemotherapy drug used to treat gastric neoplasms is epirubicin. Its applicability, treatment approach, drug type, and mechanism of action are clarified by recent research. Combination chemotherapy regimens often include epirubicin. Notably, it has been essential in perioperative settings for the treatment of resectable gastric cancer when combined with 5-fluorouracil (5-FU) and cisplatin. For the treatment of incurable gastric cancer, it has also been added to the EDP regimen along with cisplatin and docetaxel. Epirubicin is classified as a small molecule. Epirubicin is an anthracycline antineoplastic that works by attaching itself to nucleic acids. Intercalation between DNA base pairs is its main mechanism, which prevents the synthesis of both DNA and RNA. By interacting with topoisomerase II, this intercalation also causes DNA cleavage, which eventually results in cytocidal activity. Although epirubicin has shown promise, each patient will react differently. Therefore, when choosing a course of treatment, it is crucial to take into account the type of gastric neoplasm and particular patient characteristics. For individualized medical advice, speaking with a healthcare professional is advised.[1,94,108–110]

Zolbetuximab - A new medication called zolbetuximab is being used to treat gastric neoplasms. For HER2-negative gastric or gastroesophageal junction (GEJ) adenocarcinomas that are locally advanced, incurable, or metastatic and whose tumors test positive for claudin (CLDN) 18.2, zolbetuximab is the first-line treatment. It has been used in combination with chemotherapy regimens such as mFOLFOX6 and CAPOX to significantly improve overall survival (OS) and progression-free survival (PFS) in patients with advanced, HER2-negative, CLDN18.2-positive gastric/GEJ cancer. Zolbetuximab is a monoclonal antibody, which is a type of biotech medication. Up to 30% of gastric and gastroesophageal cancers overexpress the protein Claudin 18.2 (CLDN18.2), which is the target of zolbetuximab. It kills dividing cancer cells directly by binding to CLDN18.2, which also triggers the immune system to react. It causes apoptosis in gastric cancer cells that express CLDN18.2 by means of complement-dependent cytotoxicity (CDC) and antibody-dependent cellular cytotoxicity (ADCC). Zolbetuximab has demonstrated promise, but individual patient responses may differ, so it's important to take the patient's unique characteristics and the type of gastric neoplasm into account.[111]

Gimeracil - Gimeracil is a small molecule medication[1] used as an adjuvant in antineoplastic therapy. It is also referred to as 5-Chloro-2,4-dihydroxypyridine. It is sold under the brand name "Teysuno"[1] and is used in combination regimens with Tegafur and Oteracil. Gimeracil's therapeutic approach centers on chemotherapy[1], which targets the enzyme dihydropyrimidine dehydrogenase (DPD) to increase 5-Fluorouracil's (5-FU) efficacy. Gimeracil keeps 5-FU from degrading by specifically blocking DPD, guaranteeing long-term therapeutic concentrations that successfully fight cancer cells[12]. Recent studies published in 2021 showed encouraging results in terms of its suitability for treating gastric neoplasms. In patients with stage IV gastric cancer, adding Gimeracil to maintenance therapy after initial treatment resulted in a noteworthy median progression-free survival of 13.5 months and median overall survival of 23 months, as well as a high disease control rate of 87.5%.[112]

*Immunotherapy:*

Preoperative - Preoperative treatments, which use a variety of tactics described in recent literature, are being used more and more in the management of gastric neoplasms. Neoadjuvant therapy is one tactic, which entails administering medication before surgery to reduce the tumor's size and facilitate its removal[12]. Immunotherapy, which uses the body's immune system to fight cancer, is another developing tactic. To improve treatment outcomes, ongoing clinical trials are looking into immunotherapy and chemotherapy combinations for gastric or gastroesophageal junction adenocarcinomas. These treatments include chemotherapy agents like fluorouracil, leucovorin, oxaliplatin, and docetaxel, as well as biotech drugs like spartalizumab, an immunotherapy agent being studied in combination with the FLOT regimen (fluorouracil, leucovorin, oxaliplatin, and docetaxel) for the perioperative treatment of resectable gastric or gastroesophageal junction adenocarcinoma. Chemotherapy medications work by identifying and eliminating cells that divide quickly, including cancerous cells. Spartalizumab and other immunotherapy drugs work by preventing PD-1 from interacting with its ligands, PD-L1 and PD-L2, which increases T-cell activity against tumors and strengthens immune responses. Tumor characteristics, patient health, and cancer stage all play a role in the customized choice of preoperative medications and techniques. For individualized medical advice, it is imperative that patients consult healthcare professionals.[113,114]

Paclitaxel - Paclitaxel, a small molecule medication, is widely used to treat gastric neoplasms. Microtubule stabilization, which causes cell-cycle arrest and ultimately cell death, is its primary mode of action. In addition to inducing tumor cell-cycle arrest, paclitaxel has been demonstrated to enhance antitumor immunity by converting M2-polarized macrophages into the M1-like phenotype in a TLR4-dependent way. Paclitaxel's anticancer effects may be aided by this modulation of tumor-associated macrophages (TAMs) toward a more immunocompetent state through TLR4 activation, which supports its use in innovative combination therapies with immunotherapies. Paclitaxel in combination treatments has demonstrated encouraging results in recent studies. It has demonstrated strong antitumor activity in phase 1/2 trials involving patients with gastric cancer who have a variety of tumor markers, highlighting its synergy with Nivolumab. Furthermore, in subsequent treatment lines, combinations of Paclitaxel and Ramucirumab have shown positive objective response rates and progression-free survival. Although paclitaxel is a successful treatment for gastric neoplasms, individual outcomes may differ depending on the features of the tumor and general health. Therefore, when creating individualized treatment plans, healthcare providers should take these factors into account. Speaking with a healthcare expert is crucial for personalized medical advice.[115–117]

The CA19-9 antigen is a biomarker commonly used in the diagnosis and surveillance of some cancers, including gastric cancer, but it is not a therapeutic medication. Cell adhesion, motility, and immunogenicity are all impacted by this sialylated Lewis blood group antigen, which is useful for determining the state of malignant tumors and their invasive or metastatic behavior. Elevated CA19-9 levels have been seen in gastric cancer, and it may be especially helpful in identifying recurrence after a curative gastrectomy[2]. Treatment for gastric cancer usually consists of surgery, chemotherapy, and occasionally targeted therapies or immunotherapies meant to eradicate cancer cells, even though CA19-9 is essential for tracking the course of the disease and the effectiveness of treatment.[118,119]

Picibanil - Group A streptococcus is the source of picibanil, also referred to as OK-432, a formulation with anti-neoplastic qualities that is used to treat a number of illnesses, including cystic hygroma (lymphangiomas). By drawing immune cells to the injection site, the mechanism of action produces a strong inflammatory response. By using the body's immune system to fight cancer cells, this reaction stimulates T-lymphocytes, dendritic cells, and macrophages, which release a variety of cytokines and chemokines. This suggests that it may be used as an immunotherapy. Clinical trials are necessary to assess Picibanil's safety and effectiveness in treating gastric neoplasms. For individualized medical advice, always seek the advice of a healthcare professional.[120,121]

### 1.3. Objective

Recent studies suggest using the RAIN method to treat diseases by combining drugs that push the p-value linking the disease to target proteins or genes closer to one. These papers leverage AI techniques, such as Graph Neural Networks or Reinforcement Learning, to identify the best drug combinations. Typically, they conduct a network meta-analysis to assess how effective these combinations are compared to others.[122–130]

The RAIN approach is a systematic review and meta-analysis method that adheres to the STROBE guidelines to investigate a specific medical question. Unlike other recent medical studies, RAIN stands out by employing AI to tackle a particular medical issue.[131–140]

## 2. METHOD

The identification of effective drug combinations for the treatment of complex diseases presents significant challenges (see Figure 1). The vast array of potential drug combinations further complicates this process, as exhaustive experimental testing is both resource-intensive and time-consuming.

To address these challenges, we employed Graph SAGE (Graph Sample and Aggregate), a specialized Graph Neural Network (GNN) framework, to propose drug combinations likely to exhibit synergistic therapeutic effects. Graph SAGE leverages a comprehensive biomedical knowledge graph that integrates diverse data types, including drug-protein interactions, gene expression profiles, and drug-target interactions. By constructing embedding vectors to represent drugs and proteins within this knowledge graph, Graph SAGE facilitates the evaluation of similarity and structural compatibility among drug pairs. The modularity metric, which quantifies the quality of graph clustering, serves as a proxy for assessing the potential synergistic effects of drug combinations, indicating their capacity to target shared or related genes and proteins implicated in the disease.

Using Graph SAGE, we generated a curated list of drug combinations characterized by high modularity scores, suggesting their potential to effectively target disease-associated genes or proteins. However, high modularity alone does not ensure clinical efficacy or safety. To refine our findings, we applied natural language processing (NLP) techniques to analyze clinical trial data, extracting critical information such as trial outcomes, dosages, and adverse effects. This analysis enabled the exclusion of drug combinations with unfavorable or inconsistent clinical outcomes, thereby enhancing the reliability of the proposed combinations.

Subsequently, we conducted a network meta-analysis to rigorously evaluate the efficacy of the filtered drug combinations and rank them based on their effectiveness against target genes or proteins. Network meta-analysis, a robust statistical methodology, synthesizes direct and indirect evidence from multiple sources to provide a coherent and comprehensive assessment of competing interventions. Through this approach, we identified the most promising drug combinations for the treatment of the disease, accompanied by confidence intervals and ranking probabilities to support their prioritization.

In conclusion, the integration of Graph SAGE-based graph neural network analysis with NLP-driven clinical data extraction and network meta-analysis offers a powerful framework for identifying and prioritizing effective drug combinations for complex diseases. This methodology not only streamlines the selection process but also enhances the evidence-based foundation for therapeutic decision-making.

## 2.1. Stage 1 – Graph SAGE:

Graph SAGE is an inductive framework used GNNs for learning node representations, particularly in large graphs where the number of nodes and edges may be very large (see Figure 3). It is designed to effectively handle graph-structured data while providing a scalable solution for nodes that have numerous neighbors, like in the case of biomedical data such as drug-target interactions.

Key Features of Graph SAGE are as follows:

- Graph Construction: This step involves creating a graph where each node represents a drug or protein (gene). The edges represent interactions or relationships between these entities, typically with features associated with each node, such as gene expression profiles or drug properties.

- Graph SAGE Embedding Generation: Graph SAGE samples a fixed-size neighborhood of nodes for each target node. This sampling allows the model to work with large graphs by focusing on a smaller, fixed-size subset of neighbors. The neighborhood is then aggregated using a differentiable aggregation function (mean, LSTM, or pooling). This method allows the aggregation of local structural information. Formula for the node embedding update:

$$h_v^{(k)} = Agg(\{h_u^{(k-1)} \mid u \in N(v)\}) \parallel h_v^{(k-1)}$$

where $h_v^{(k)}$ is the embedding of node $v$ at the $k^{th}$ layer, N(v) is the set of neighbors of node $v$, Agg is the aggregation function (e.g., mean, LSTM, or pooling), and ∥ denotes concatenation of features from the node itself and its neighbors.

- Drug Combination Prediction: After generating embeddings for nodes (drugs or genes), these embeddings are compared to predict potential drug combinations. The similarity between drug embeddings is evaluated using distance metrics like cosine similarity.

- Validation and Refinement: Clinical trial data is used to validate predicted medication combinations. The predictions for the best drug combinations are improved by comparing the efficacy of different combinations using a network meta-analysis (NMA) based on clinical trial results. The goal is to evaluate both the therapeutic impact and potential molecular interactions between drugs and disease-related genes or proteins.

Graph SAGE offers a powerful way to incorporate both biological information (e.g. gene expression) and structural data (such as drug interactions) to predict new and effective drug combinations for diseases like gastric neoplasm, improving treatment outcomes.

Our method offers a scalable and reliable way to recommend drug combinations by utilizing GraphSAGE within this framework. In the end, this methodology supports researchers and healthcare professionals in their efforts to improve patient outcomes by providing insights into the disease's underlying mechanisms in addition to helping identify effective treatments.[141–143]

## 2.2. Stage 2- NLP for Article Retrieval:

This systematic review and meta-analysis looked for relevant studies using the keywords prevalence, epidemiology, vasovagal syncope, and reflex syncope. Searches were conducted using a variety of databases, including PubMed, WoS, Scopus, ScienceDirect, and Google Scholar, with no time restrictions until July 20, 2022. In order to increase the number of related studies, references from identified articles were manually reviewed after the extracted data was arranged using EndNote software.

**Inclusion and Exclusion Criteria:** The studies had to meet the following requirements in order to be included: 1) to be cross-sectional studies; 2) have the whole text accessible; 3) have sufficient data, including sample size and prevalence rates; and 4) examine the prevalence of vasovagal syncope in various countries. The exclusion criteria included duplicate research, conference papers, case reports, and case series.

**Study Choosing:** The studies were managed using EndNote, and duplicates were removed first. Titles and abstracts were examined during the first screening, and studies that were deemed unnecessary were eliminated. Studies that satisfied the inclusion criteria were kept after the full texts were examined during the secondary evaluation stage.

**Quality Assessment:** The Strengthening the Reporting of Observational Studies in Epidemiology (STROBE) checklist was used to evaluate the studies' quality. This checklist covers 32 items pertaining to the study's methodology and assesses six domains: title, abstract, introduction, methods, results, and discussion. Articles scoring 16 or higher were considered to be of good or moderate quality, while those scoring lower than 16 were eliminated due to poor methodology.

**Data Extraction:** Two researchers extracted data using a pre-made checklist that included details on the author, publication year, study location, sample size, participant age and gender, data collecting techniques, and prevalence rates.

**Statistical Analysis:** The I2 test was used to measure study heterogeneity, and because of the significant heterogeneity, the results were synthesized using a random effects model. Egger's test was used to assess publication bias, and funnel plots were produced. Comprehensive Meta-Analysis software (version 2) was used to analyze the data, and meta-regression analysis was done to investigate the sample size and publication year—two factors that contribute to heterogeneity.

According to PRISMA guidelines, as it is shown in Figure 2, the systematic review and meta-analysis shed light on the prevalence of vasovagal syncope worldwide. Duplicate articles were eliminated out of articles that were found in databases.[144]

## 2.3. Stage 3 - Network Meta-Analysis for Effectiveness Assessment:

By combining known drug-disease associations and drug-drug and disease-disease similarities, the NMA framework creates an undirected graph that represents a heterogeneous network. It uses meta paths of varying lengths to capture indirect relationships. NMA creates low-dimensional representation vectors that encode these relationships by analyzing the network. The learned representations and indirect links are then used by a classifier to predict new drug-disease associations.[145,146]

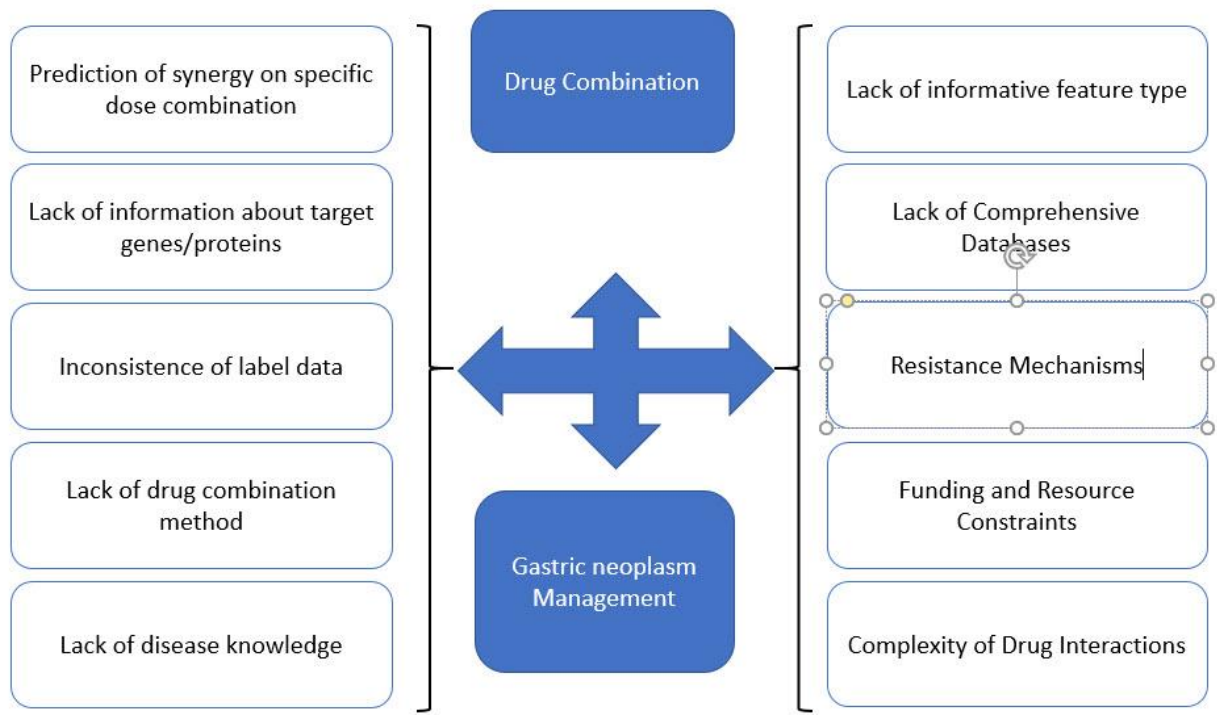

Figure 1: How suggested medication combinations affect the handling of cases of gastric neoplasms.

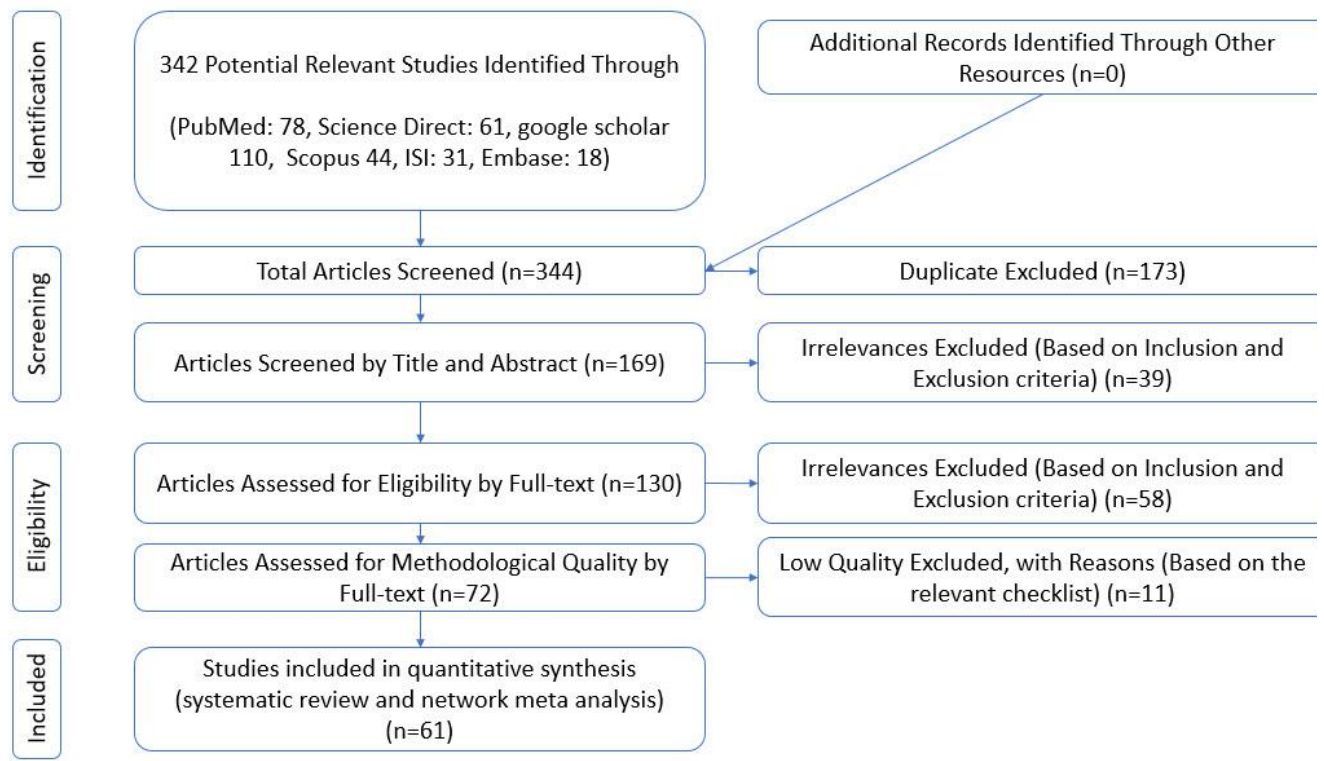

Figure 2: A PRISMA (2020) flow diagram showing the steps involved in the RAIN method's article sieving

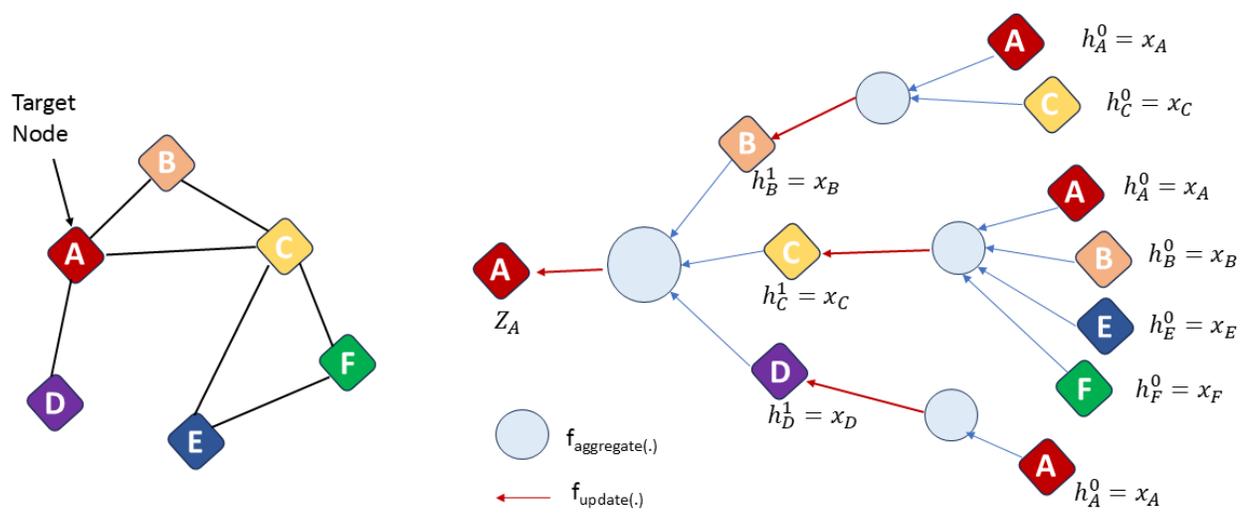

Figure 3: suggested approach

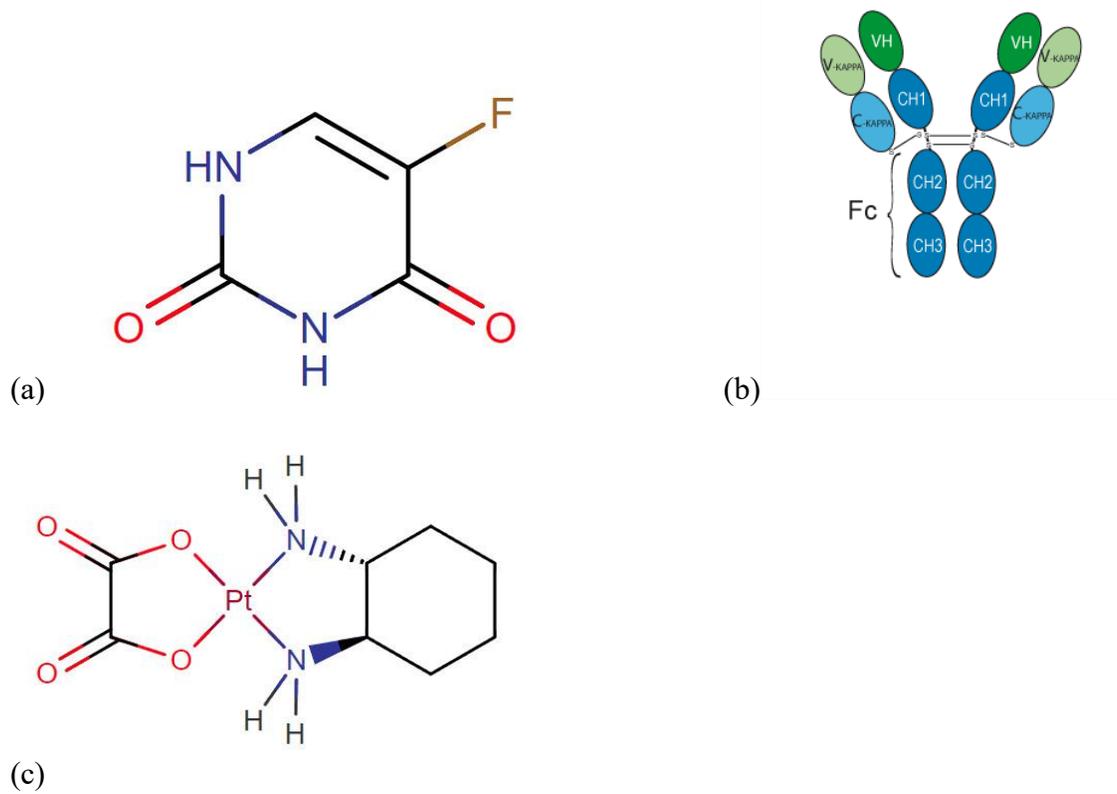

(a)

(b)

(c)

Figure 4: Drug structure for (a) Fluorouracil, (b) Trastuzumab, (c) Oxaliplatin from https://www.drugbank.com/

Table 1: p-value between scenarios and Gastric neoplasm

| scenario | drug combination | p-value |
|---|---|---|
| s1 | Fluorouracil | 0.02291194 |
| s2 | s1 + Trastuzumab | 0.009918244 |
| s3 | s2 + Oxaliplatin | 0.006902485 |

Table 2: p-value between Gastric neoplasm and human proteins/genes after implementing scenarios

| Association Name | S0 | S1 | S2 | S3 | Association Name | S0 | S1 | S2 | S3 |
|---|---|---|---|---|---|---|---|---|---|
| CDH1 | 2/6E-05 | 0/99738 | 0/99998 | 1 | CDH2 | 0/00033 | 0/98926 | 0/99788 | 0/99997 |
| FZR1 | 2/6E-05 | 0/60798 | 0/99729 | 0/99967 | MSH2 | 0/00034 | 0/99681 | 0/99936 | 1 |
| ERBB2 | 2/9E-05 | 0/99986 | 1 | 1 | TYMP | 0/00034 | 0/99999 | 1 | 1 |
| MUC6 | 5/9E-05 | 5/9E-05 | 0/83757 | 0/83757 | VEGFA | 0/00034 | 0/99934 | 1 | 1 |
| GAST | 6/1E-05 | 0/87694 | 0/93827 | 0/97122 | CD44 | 0/00035 | 0/9983 | 0/99999 | 1 |
| TP53 | 8/5E-05 | 0/99985 | 1 | 1 | CHGA | 0/00037 | 0/94844 | 0/94844 | 0/94844 |
| CLDN18 | 8/5E-05 | 0/79928 | 0/9985 | 0/99999 | RNF180 | 0/00037 | 0/93428 | 0/93428 | 0/93428 |
| GKN1 | 0/00011 | 0/968 | 0/968 | 0/9978 | GACAT3 | 0/00037 | 0/00037 | 0/00037 | 0/00037 |
| CDX2 | 0/00012 | 0/98205 | 0/99406 | 0/99994 | CDKN1A | 0/00038 | 0/99939 | 0/99993 | 1 |
| MUC5AC | 0/00013 | 0/69212 | 0/69212 | 0/94687 | COMMD6 | 0/00039 | 0/97585 | 0/99946 | 0/99946 |
| RUNX3 | 0/00013 | 0/97953 | 0/97953 | 0/99755 | FGFR2 | 0/0004 | 0/90444 | 0/99921 | 0/99998 |
| CD274 | 0/00014 | 0/99677 | 1 | 1 | VIM | 0/0004 | 0/98849 | 0/98849 | 0/99989 |
| MLH1 | 0/00014 | 0/99853 | 0/99973 | 1 | CEACAM5 | 0/00041 | 0/00041 | 0/00041 | 0/00041 |
| MET | 0/00014 | 0/97821 | 0/99997 | 1 | MYC | 0/00042 | 0/99371 | 0/99951 | 0/99999 |
| TSC1 | 0/00014 | 0/99604 | 0/99968 | 0/99999 | MMP7 | 0/00043 | 0/97909 | 0/98959 | 0/99888 |
| MUC2 | 0/00015 | 0/96922 | 0/98146 | 0/9877 | ZNF135 | 0/00045 | 0/99728 | 0/99976 | 1 |
| GKN2 | 0/00018 | 0/00018 | 0/00018 | 0/97384 | STAT3 | 0/00045 | 0/99366 | 0/99993 | 1 |
| LOC100508689 | 0/00018 | 0/00018 | 0/00018 | 0/00018 | RPRM | 0/00046 | 0/76355 | 0/76355 | 0/76355 |
| EGFR | 0/00019 | 0/99969 | 1 | 1 | H3F3AP6 | 0/00047 | 0/00047 | 0/00047 | 0/00047 |
| BCL2 | 0/0002 | 0/99972 | 0/99999 | 1 | TCEAL1 | 0/00047 | 0/99929 | 0/99989 | 1 |
| TFF2 | 0/00021 | 0/00021 | 0/00021 | 0/00021 | PDCD1 | 0/00048 | 0/00048 | 0/00048 | 0/00048 |
| PSCA | 0/00021 | 0/00021 | 0/93064 | 0/9736 | AFP | 0/00048 | 0/00048 | 0/00048 | 0/00048 |
| TFF1 | 0/00023 | 0/00023 | 0/96145 | 0/96145 | MMP9 | 0/0005 | 0/0005 | 0/0005 | 0/0005 |
| ARID1A | 0/00026 | 0/06248 | 0/98949 | 0/99979 | CASP3 | 0/0005 | 0/0005 | 0/0005 | 0/0005 |
| S100A8 | 0/00028 | 0/93121 | 0/93121 | 0/93121 | ATP4A | 0/00051 | 0/00051 | 0/00051 | 0/00051 |
| PGC | 0/0003 | 0/0003 | 0/0003 | 0/0003 | PTEN | 0/00053 | 0/00053 | 0/00053 | 0/00053 |
| DPYD | 0/0003 | 1 | 1 | 1 | ERCC1 | 0/00054 | 0/99988 | 1 | 1 |
| TYMS | 0/0003 | 1 | 1 | 1 | CCKBR | 0/00056 | 0/68252 | 0/68252 | 0/9464 |
| CCND1 | 0/00031 | 0/99797 | 0/99999 | 1 | KIT | 0/00056 | 0/99313 | 0/9999 | 0/9999 |
| CTNNB1 | 0/00032 | 0/96269 | 0/99683 | 0/99994 | MMP2 | 0/00056 | 0/96447 | 0/96447 | 0/99555 |

Table 3: characteristics of suggested medications as efficient treatments for gastric neoplasms.

| Drug name | Accession Number | Type - Small Molecule | Type - Biotech | Formula | Mechanism of Action |
|---|---|---|---|---|---|
| Fluorouracil | DB00544 | * | | $C_4H_3FN_2O_2$ | The exact mechanism of fluorouracil remains incompletely understood, but it is primarily believed to involve the formation of a stable ternary complex between the drug's deoxyribonucleotide (FdUMP), the folate cofactor N5–10-methylenetetrahydrofolate, and thymidylate synthase (TS). This complex effectively inhibits the conversion of uracil to thymidylate, leading to disruptions in DNA and RNA synthesis and subsequent cell death. Additionally, fluorouracil can be mistakenly integrated into RNA in place of uridine triphosphate (UTP), resulting in defective RNA that impairs RNA processing and protein production. |
| Trastuzumab | DB00072 | | * | $C_{6470}H_{10012}N_{1726}O_{2013}S_{42}$ | Trastuzumab is a recombinant humanized IgG1 monoclonal antibody that targets the HER-2 receptor, part of the epidermal growth factor receptor (EGFR) family, which is frequently overexpressed in breast tumor cells. HER-2 amplifies signals from other HER family receptors by forming heterodimers, leading to activation of downstream signaling pathways like Ras/Raf/MAPK, PI3K/Akt, and PLCγ/PKC, promoting cell growth and survival. Trastuzumab binds to the extracellular domain of HER-2, blocking its cleavage and downmodulating receptor activity, which inhibits these signaling pathways and induces cell cycle arrest. It also enhances immune response through antibody-dependent cell-mediated cytotoxicity (ADCC) by attracting immune cells to HER-2-overexpressing tumor sites. While trastuzumab has limited potential to induce complement-dependent cytotoxicity (CDC) on its own, its combination with pertuzumab can activate the complement system, showing synergistic effects in cancer treatment. However, some patients with HER-2 positive breast cancer exhibit intrinsic resistance to trastuzumab, often due to alterations in phosphatase and tensin homologue (PTEN) or overactivation of phosphoinositide 3-kinase, as well as the overexpression of other receptors like the insulin-like growth factor receptor. |
| Oxaliplatin | DB00526 | * | | $C_8H_{14}N_2O_4Pt$ | Oxaliplatin undergoes nonenzymatic conversion in physiological environments, resulting in the formation of active derivatives through the displacement of its oxalate ligand. This process generates several transient reactive species, including monoaquo and diaquo DACH platinum, which form covalent bonds with macromolecules. These bonds create both interstrand and intrastrand platinum-DNA crosslinks, particularly between the N7 positions of adjacent guanines (GG), adjacent adenine-guanines (AG), and guanines separated by an intervening nucleotide (GNG). These crosslinks disrupt DNA replication and transcription, leading to cytotoxic effects that are not unique to any one stage of the cell cycle. |

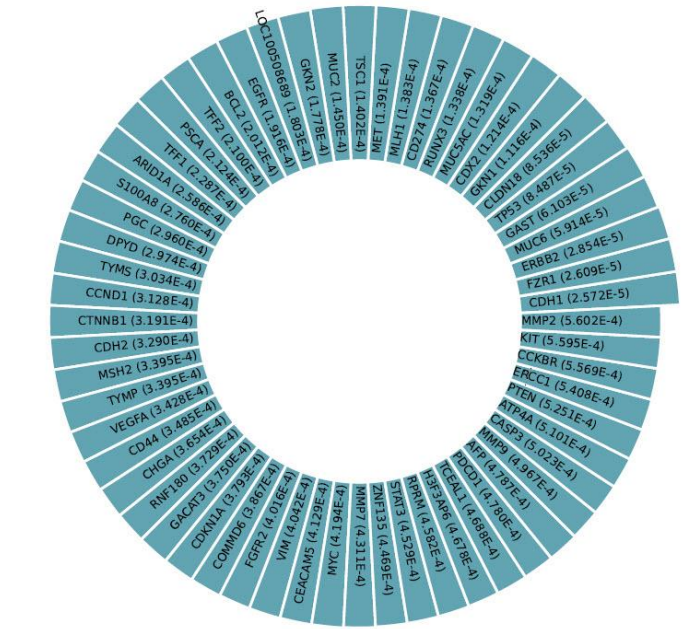

Figure 5: p-values between affected human proteins/genes and Gastric neoplasm

## 3. RESULTS

### 3.1. Stage 1- Graph SAGE:

Fluorouracil Trastuzumab and oxaliplatin are among the medication combinations that the GNN recommends. Figure 4 illustrates their structures and Table 1 details the pairing's significance and displays the corresponding p values. For instance, fluorouracil and gastric neoplasms (Scenario 1) have a p value of 0.0229. But when Trastuzumab is added (Scenario 2), this value drops dramatically to 0.0099. Additionally, the p-value in the third scenario is 0.0069, demonstrating the beneficial impact of this medication combination on disease control. The difference in p-values between human proteins/genes and gastric neoplasms under different circumstances is shown in Table 2. The p-values between gastric neoplasms and the corresponding impacted human proteins or genes are displayed in Column "S0". Table 3 elucidates each drug's characteristics.

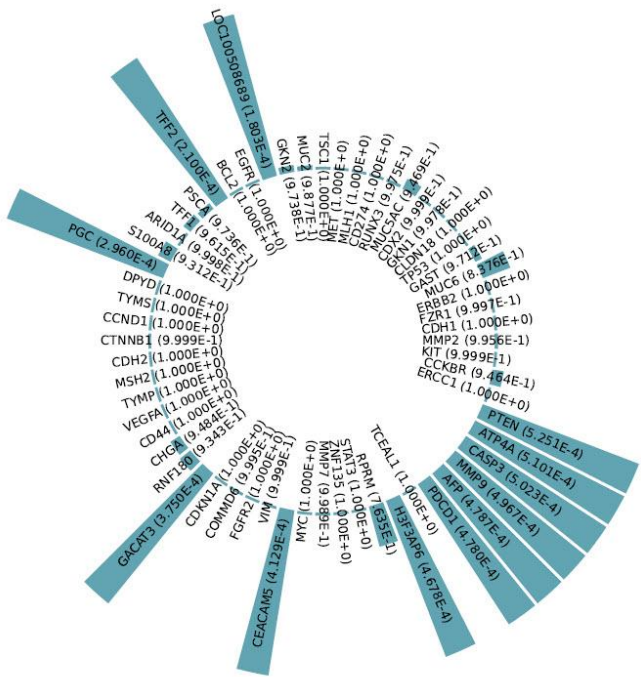

Figure 6: p-values between affected human proteins/genes and Gastric neoplasm after implementing scenario 4.

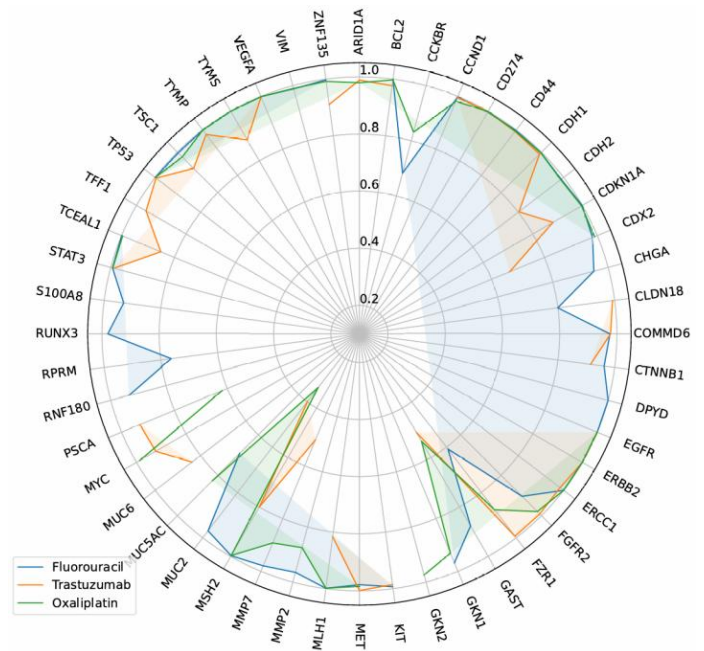

Figure 7: radar-plot for NMA

### 3.2. Stage 2- NLP for Article Retrieval:

In order to find relevant studies for this systematic review and meta-analysis, researchers used predefined keywords to search several databases, including PubMed, Web of Science (WoS), Scopus, Science Direct, and Google Scholar. The collected data was arranged using Python software. Included were studies that examined the prevalence of gastric neoplasm in various nations; Preferable were cross-sectional studies with thorough texts and sufficient data. The analysis did not include conference articles, case reports, case series, or replication studies. The selection method carefully reviewed titles, abstracts, and full texts to exclude duplicates and research that did not meet the inclusion requirements. Sixty-one articles in all were upgraded for qualitative assessment. All review procedures, including study selection, quality evaluation, and data extraction, were carried out by two independent investigators to minimize bias. Any disagreements were settled by discussion or consultation with a third investigator.

Table 4: significant studies for suggested medications in the treatment of gastric neoplasms

| study name | Fluorouracil | Trastuzumab | Oxaliplatin | study name | Fluorouracil | Trastuzumab | Oxaliplatin |
|---|---|---|---|---|---|---|---|
| **Shi Y. , 2024** | * | * |   | Hofheinz RD. , 2022 | * | * | * |
| **Nakamura A. , 2024** | * |   | * | Aktürk Esen S. , 2022 | * | * | * |
| **Arias-Martinez A. , 2024** | * |   | * | Nishie N. , 2022 | * | * | * |
| **Hirose T. , 2024** | * |   | * | Zhang Q. , 2022 | * | * | * |
| **Bir Yücel K. , 2024** | * |   | * | Schade S. , 2022 | * | * | * |
| **Zhao Y. , 2024** | * |   | * | Ladak I. , 2022 | * | * |   |
| **Qiu MZ. , 2024** | * |   | * | Hofheinz RD. , 2022 | * | * |   |
| **Egebjerg K. , 2024** | * |   | * | Wu XQ. , 2022 | * | * |   |
| **Zhao S. , 2024** | * |   | * | Aktürk Esen S. , 2022 | * | * |   |
| **Che G. , 2024** | * |   | * | Nishie N. , 2022 | * | * |   |
| **Martins D. , 2024** | * |   | * | Zhang Q. , 2022 | * | * |   |
| **Tantia P. , 2024** | * |   | * | Schade S. , 2022 | * | * |   |
| **Salla M. , 2024** | * |   | * | Mori Y. , 2022 |   | * | * |
| **Ntwali F. , 2024** | * |   | * | Hofheinz RD. , 2022 |   | * | * |
| **Mineur L. , 2024** | * |   | * | Aktürk Esen S. , 2022 |   | * | * |
| **Li N. , 2024** | * |   | * | Hofheinz RD. , 2021 | * | * | * |
| **Li N. , 2024** |   | * | * | Grieb BC. , 2021 | * | * | * |
| **Wakatsuki T. , 2024** |   | * | * | Horinouchi T. , 2021 | * | * | * |
| **Yanagisawa N. , 2024** |   | * | * | Jimenez-Fonseca P. , 2021 | * | * | * |
| **Hirase Y. , 2024** |   | * | * | Hofheinz RD. , 2021 | * | * |   |
| **Zhou JH. , 2024** |   | * | * | Grieb BC. , 2021 | * | * |   |
| **Yamane H. , 2024** |   | * | * | Yamashita K. , 2021 | * | * |   |
| **Fuchino M. , 2024** |   | * | * | Horinouchi T. , 2021 | * | * |   |
| **Liu Y. , 2023** | * | * |   | Jimenez-Fonseca P. , 2021 | * | * |   |
| **Yoon J. , 2023** | * | * |   | Al-Batran SE. , 2020 | * | * | * |
| **Shirasaki Y. , 2023** |   | * | * | Park H. , 2020 | * | * | * |
| **Shimizu K. , 2023** |   | * | * | Janjigian YY. , 2020 | * | * | * |
| **Li S. , 2023** |   | * | * | Tintelnot J. , 2020 | * | * | * |
| **Iwamuro M. , 2023** |   | * | * | Catenacci DVT. , 2020 | * | * | * |
| **Huo B. , 2023** |   | * | * | Gutting T. , 2019 | * | * | * |

The Strengthening Reporting of Observational Studies in Epidemiology (STROBE) guidelines, which placed an emphasis on a number of methodological criteria, were followed in the quality assessment process. Articles with a score of 16 or higher—which were considered to be of moderate or high quality—were included in the final analysis.

As part of the data extraction process, important facts from the selected studies were noted, such as the author's identity, the year of publication, the study location, sample size, participant demographics, data collection methods, and prevalence rates. The random effects model was used for statistical analysis because of the high degree of variability among the studies. Throughout the research process, the final systematic review and meta-analysis followed PRISMA guidelines and offered insights into the worldwide prevalence of gastric neoplasm.

### 3.3. Stage 3 - Interpretation and Network meta-analysis

Figure 5 shows the p-values for human proteins and genes impacted by gastric neoplasia, and Figure 6 shows the p-values when the third scenario was applied. A radar chart in Figure 7, which displays the p-values between human proteins and genes and gastric tumor after the administration of the chosen treatments, is used to demonstrate the efficacy of the medications identified by the drug selection algorithm. Figure 8 displays the p-values illustrating the connections between targets and various interface components. The colors blue and green below represent P-values. 01 and .05, in that order. The effectiveness of the particular medication in that particular circumstance is indicated by each line of various colors.

Table 4 lists some significant studies for suggested medications in the treatment of gastric neoplasms.[93,147–190]

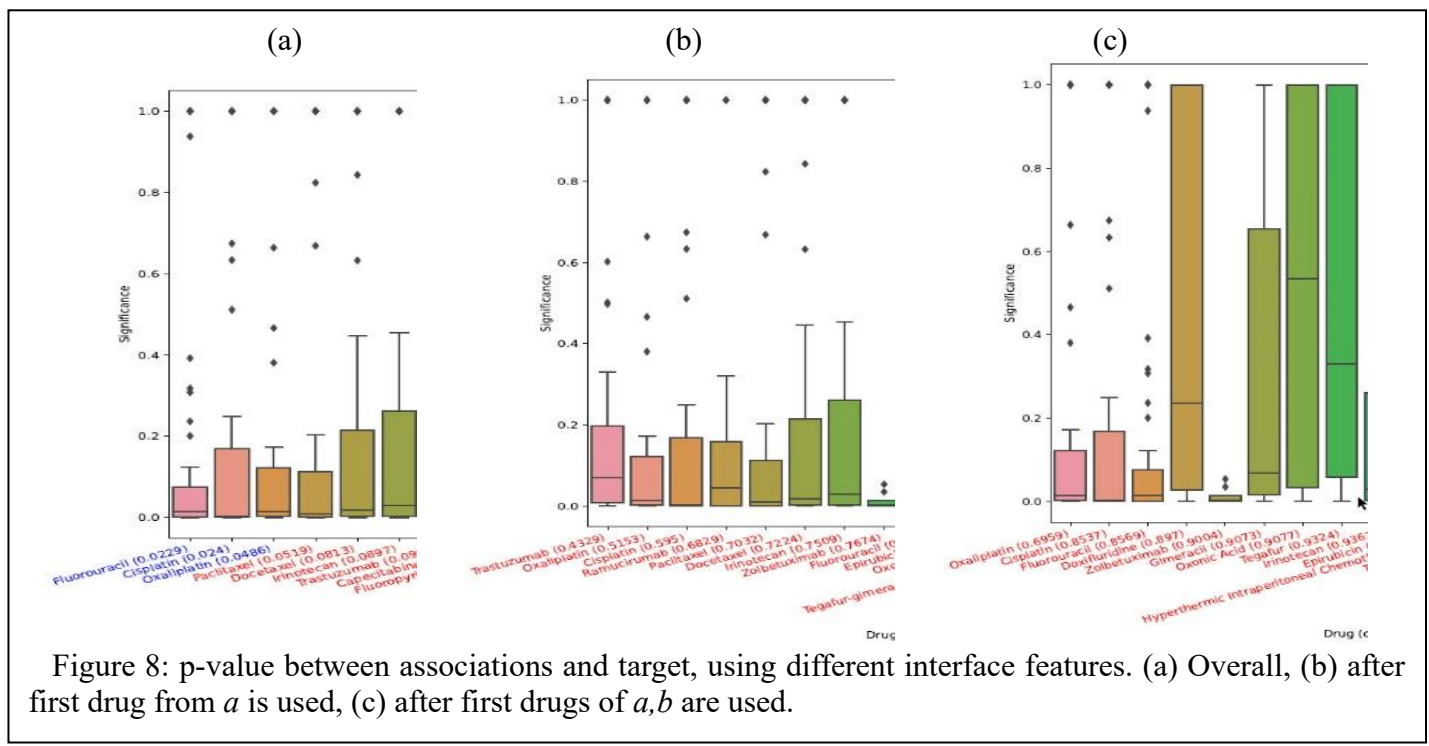

Figure 8: p-value between associations and target, using different interface features. (a) Overall, (b) after first drug from *a* is used, (c) after first drugs of *a,b* are used.

## 4. DISCUSSION

Neutropenia can result from trastuzumab treatment alone, but this effect can be intensified when combined with other immunosuppressive medications. According to clinical trials, patients who receive both myelosuppressive chemotherapy and trastuzumab have greater rates of febrile neutropenia and moderate to severe neutropenia than patients who only receive chemotherapy. The underlying reason for this worsened neutropenia is still unknown, though. Furthermore, combining oxaliplatin with other QT-prolonging medications may increase the risk of additional QT prolongation because it is known to cause ventricular arrhythmias and lengthen the QT interval.

## 5. CONCLUSION

This study underscores the potential of artificial intelligence, particularly the RAIN method, in enhancing the treatment strategies for gastric neoplasms. By combining the Graph SAGE model with network meta-analysis, we successfully identified promising drug combinations that demonstrate improved therapeutic efficacy in gastric cancer treatment. Fluorouracil, trastuzumab, and oxaliplatin emerged as highly effective drug combinations, offering valuable insights into optimizing treatment regimens.

The use of AI-driven decision-making tools, such as the RAIN method, empowers clinicians and health policymakers to make more informed, data-driven decisions in selecting drug combinations. This approach helps clarify the molecular mechanisms underlying gastric cancer and opens up new avenues for targeted therapies. Through the integration of clinical trial data, molecular insights, and drug interaction networks, the RAIN method presents a robust strategy for improving patient outcomes.

Ultimately, the findings from this research demonstrate how artificial intelligence can be applied to solve complex medical challenges, offering a path toward personalized, more effective treatments for gastric neoplasms and other cancers. In addition to helping find possible drug combinations, the method advances our knowledge of the genetic and molecular elements that influence the course of disease, allowing for more targeted and customized treatment approaches.

*Abbreviations*
STROBE: enhancing the reporting of epidemiological observational studies; PRISMA stands for Preferred Reporting Items for Meta-Analysis and Systematic Reviews. RAIN: Systematic Review and Meta-Analysis of Artificial Intelligence Networks

*Authors' contributions*
**SZ Pirasteh:** Collecting information and creating an initial draft. **Ali A. Kiaei:** Methodology, conceptualization, review and editing, formal analysis, supervision, and the use of AI models. **Mahnaz Bush:** Methodology, AI model implementation, and conceptualization. **Raha Aghaei**: ideation, research, editing and review, and verification. **Behnaz Sadeghigol:** Review & Editing. **Sabra Moghadam:** Investigation, Investigation, Review & Editing.

*Disclosure of AI Usage*
This work utilized an artificial intelligence tool for paraphrasing certain sections of the text to enhance clarity and readability. All paraphrased content has been carefully reviewed to ensure originality and proper attribution to existing sources. The intellectual contributions and core ideas remain the sole work of the authors.

*Funding: Not applicable.*

*Availability of data and materials:* **Datasets are** available through the corresponding author upon reasonable request.

*Ethics approval and consent to participate: Not applicable.*

*Consent for publication*: **Not applicable.**

*Conflict of interests:* **The authors have no conflict of interest.**

# REFERENCES:

1. Guan, W.-L., He, Y. & Xu, R.-H. Gastric cancer treatment: recent progress and future perspectives. *J. Hematol. Oncol.J Hematol Oncol* **16**, 57 (2023).
2. Panda, S. K. *et al.* Evolution of treatment in gastric cancer- a systematic review. *J. Egypt. Natl. Cancer Inst.* **34**, 1–5 (2022).
3. Cobani, E. *et al.* Gastric Cancer Survivorship: Multidisciplinary Management, Best Practices and Opportunities. *J. Gastrointest. Cancer* **55**, 519–533 (2024).
4. Mullen, J. T. Top Gastric Cancer Articles from 2022 and 2023 to Inform Your Cancer Practice. *Ann. Surg. Oncol.* **31**, 3978–3983 (2024).
5. Xie, D. *et al.* The Potential Role of CDH1 as an Oncogene Combined With Related miRNAs and Their Diagnostic Value in Breast Cancer. *Front. Endocrinol.* **13**, (2022).
6. Gregory, S. N. & Davis, J. L. CDH1 and hereditary diffuse gastric cancer: a narrative review. *Chin. Clin. Oncol.* **12**, (2023).
7. Crawford, L. J., Anderson, G., Johnston, C. K. & Irvine, A. E. Identification of the APC/C co-factor FZR1 as a novel therapeutic target for multiple myeloma. *Oncotarget* **7**, 70481–70493 (2016).
8. Subramanian, J., Katta, A., Masood, A., Vudem, D. R. & Kancha, R. K. Emergence of ERBB2 Mutation as a Biomarker and an Actionable Target in Solid Cancers. *The Oncologist* **24**, e1303–e1314 (2019).
9. Wang, S. *et al.* ERBB2D16 Expression in HER2 Positive Gastric Cancer Is Associated With Resistance to Trastuzumab. *Front. Oncol.* **12**, (2022).
10. Liu, B.-H. *et al.* Tumor Suppressive Role of MUC6 in Wilms Tumor via Autophagy-Dependent β-Catenin Degradation. *Front. Oncol.* **12**, (2022).
11. Miao, Z.-F. *et al.* Progress and remaining challenges in comprehensive gastric cancer treatment. *Holist. Integr. Oncol.* **1**, 4 (2022).
12. Chen, C., Yang, Y., Li, P. & Hu, H. Incidence of Gastric Neoplasms Arising from Autoimmune Metaplastic Atrophic Gastritis: A Systematic Review and Case Reports. *J. Clin. Med.* **12**, 1062 (2023).
13. Wu, C.-E. *et al.* p53 as a biomarker and potential target in gastrointestinal stromal tumors. *Front. Oncol.* **12**, (2022).
14. Su, Y. *et al.* Current insights into the regulation of programmed cell death by TP53 mutation in cancer. *Front. Oncol.* **12**, (2022).
15. Wang, C. *et al.* CLDN18.2 expression and its impact on prognosis and the immune microenvironment in gastric cancer. *BMC Gastroenterol.* **23**, 1–14 (2023).
16. Liu, S. *et al.* Claudin-18.2 mediated interaction of gastric Cancer cells and Cancer-associated fibroblasts drives tumor progression. *Cell Commun. Signal.* **22**, 1–16 (2024).
17. Recent Advances in Gastrointestinal Neuroendocrine Tumor Classification. https://journalononcology.org/articles/joo-v4-1134.html.
18. Koper-Lenkiewicz, O. M., Kamińska, J., Gawrońska, B. & Matowicka-Karna, J. <p>The role and diagnostic potential of gastrokine 1 in gastric cancer</p>. *Cancer Manag. Res.* **11**, 1921–1931 (2019).
19. Halder, A. *et al.* CDX2 expression in gastric carcinoma: A clinicopathological study. *Indian J. Med. Paediatr. Oncol.* **39**, 52–57 (2018).
20. Krishn, S. R., Ganguly, K., Kaur, S. & Batra, S. K. Ramifications of secreted mucin MUC5AC in malignant journey: a holistic view. *Carcinogenesis* **39**, 633–651 (2018).
21. Suda, K. *et al.* Aberrant Upregulation of RUNX3 Activates Developmental Genes to Drive Metastasis in Gastric Cancer. *Cancer Res. Commun.* **4**, 279–292 (2024).

22. Wei, D. *et al.* Loss of RUNX3 Expression Significantly Affects the Clinical Outcome of Gastric Cancer Patients and Its Restoration Causes Drastic Suppression of Tumor Growth and Metastasis. *Cancer Res.* **65**, 4809–4816 (2005).
23. Zhou, X. *et al.* Integrative study reveals the prognostic and immunotherapeutic value of CD274 and PDCD1LG2 in pan-cancer. *Front. Genet.* **13**, (2022).
24. Wang, Y. *et al.* Prognostic value and immunological role of PD-L1 gene in pan-cancer. *BMC Cancer* **24**, 1–19 (2024).
25. Corso, G. Microsatellite instability in gastrointestinal cancers. *Eur. J. Hum. Genet.* **30**, 996–997 (2022).
26. Plotkin, A., Olkhov-Mitsel, E. & Nofech-Mozes, S. MLH1 Methylation Testing as an Integral Component of Universal Endometrial Cancer Screening—A Critical Appraisal. *Cancers* **15**, 5188 (2023).
27. Reddavid, R. *et al.* Molecularly Targeted Therapies for Gastric Cancer. State of the Art. *Cancers* **13**, 4094 (2021).
28. Mallela, K. & Kumar, A. Role of TSC1 in physiology and diseases. *Mol. Cell. Biochem.* **476**, 2269–2282 (2021).
29. Conze, T. *et al.* MUC2 mucin is a major carrier of the cancer-associated sialyl-Tn antigen in intestinal metaplasia and gastric carcinomas. *Glycobiology* **20**, 199–206 (2010).
30. Baccili Cury Megid, T., Farooq, A. R., Wang, X. & Elimova, E. Gastric Cancer: Molecular Mechanisms, Novel Targets, and Immunotherapies: From Bench to Clinical Therapeutics. *Cancers* **15**, 5075 (2023).
31. Zhang, Z. *et al.* GKN2 promotes oxidative stress-induced gastric cancer cell apoptosis via the Hsc70 pathway. *J. Exp. Clin. Cancer Res.* **38**, 338 (2019).
32. Corso, S. *et al.* Optimized EGFR Blockade Strategies in EGFR Addicted Gastroesophageal Adenocarcinomas. *Clin. Cancer Res.* **27**, 3126–3140 (2021).
33. Ding, B. *et al.* EGFR and MMP7 are important targets for gastric cancer metastasis. Preprint at https://doi.org/10.21203/rs.3.rs-3604422/v1 (2023).
34. Scarfò, L. Novel therapies and combinations in CLL refractory to BTK inhibitors and venetoclax. *Hematology* **2022**, 316–322 (2022).
35. Kaloni, D., Diepstraten, S. T., Strasser, A. & Kelly, G. L. BCL-2 protein family: attractive targets for cancer therapy. *Apoptosis* **28**, 20–38 (2023).
36. Qian, Z. *et al.* Validation of the DNA Methylation Landscape of TFF1/TFF2 in Gastric Cancer. *Cancers* **14**, 5474 (2022).
37. Nayerpour Dizaj, T. *et al.* Significance of PSCA as a novel prognostic marker and therapeutic target for cancer. *Cancer Cell Int.* **24**, 135 (2024).
38. Wu, D. *et al.* PSCA is a target of chimeric antigen receptor T cells in gastric cancer. *Biomark. Res.* **8**, 3 (2020).
39. Hoffmann, W. Self-Renewal and Cancers of the Gastric Epithelium: An Update and the Role of the Lectin TFF1 as an Antral Tumor Suppressor. *Int. J. Mol. Sci.* **23**, 5377 (2022).
40. Lu, S., Duan, R., Cong, L. & Song, Y. The effects of ARID1A mutation in gastric cancer and its significance for treatment. *Cancer Cell Int.* **23**, 296 (2023).
41. Angelico, G. *et al.* ARID1A Mutations in Gastric Cancer: A Review with Focus on Clinicopathological Features, Molecular Background and Diagnostic Interpretation. *Cancers* **16**, 2062 (2024).
42. Zhou, H., Zhao, C., Shao, R., Xu, Y. & Zhao, W. The functions and regulatory pathways of S100A8/A9 and its receptors in cancers. *Front. Pharmacol.* **14**, (2023).
43. Lin, Q. *et al.* S100A8 is a prognostic signature and associated with immune response in diffuse large B-cell lymphoma. *Front. Oncol.* **14**, (2024).
44. Lv, H., Zhou, D. & Liu, G. PVT1/miR-16/CCND1 axis regulates gastric cancer progression. *Open Med.* **18**, (2023).

45. Cai, N. *et al.* Targeting MMP9 in CTNNB1 mutant hepatocellular carcinoma restores CD8+ T cell-mediated antitumour immunity and improves anti-PD-1 efficacy. *Gut* **73**, 985–999 (2024).
46. Parker, J., Hockney, S., Blaschuk, O. W. & Pal, D. Targeting N-cadherin (CDH2) and the malignant bone marrow microenvironment in acute leukaemia. *Expert Rev. Mol. Med.* **25**, e16 (2023).
47. Lim, H. J., Zhuang, L. & Fitzgerald, R. C. Current advances in understanding the molecular profile of hereditary diffuse gastric cancer and its clinical implications. *J. Exp. Clin. Cancer Res.* **42**, 57 (2023).
48. Evolving Landscape of Targeted Treatment Options for HER2-Positive Gastric/Gastroesophageal Adenocarcinomas – Hematology & Oncology. https://www.hematologyandoncology.net/archives/june-2023/evolving-landscape-of-targeted-treatment-options-for-her2-positive-gastric-gastroesophageal-adenocarcinomas/.
49. Paladhi, A., Daripa, S., Mondal, I. & Hira, S. K. Targeting thymidine phosphorylase alleviates resistance to dendritic cell immunotherapy in colorectal cancer and promotes antitumor immunity. *Front. Immunol.* **13**, (2022).
50. Liao, C. *et al.* CD44 Glycosylation as a Therapeutic Target in Oncology. *Front. Oncol.* **12**, (2022).
51. Gómez-Gallegos, A. A. *et al.* CD24+CD44+CD54+EpCAM+ gastric cancer stem cells predict tumor progression and metastasis: clinical and experimental evidence. *Stem Cell Res. Ther.* **14**, 16 (2023).
52. Hou, W., Kong, L., Hou, Z. & Ji, H. CD44 is a prognostic biomarker and correlated with immune infiltrates in gastric cancer. *BMC Med. Genomics* **15**, 225 (2022).
53. Wang, T., Zhang, Y., Wang, J. & Li, Y. Diagnostic value of plasma RNF180 gene methylation for gastric cancer: A systematic review and meta-analysis. *Front. Oncol.* **12**, (2023).
54. Deng, J. *et al.* Clinical and experimental role of ring finger protein 180 on lymph node metastasis and survival in gastric cancer. *Br. J. Surg.* **103**, 407–416 (2016).
55. Yuan, X., Dong, Z. & Shen, S. LncRNA GACAT3: A Promising Biomarker and Therapeutic Target in Human Cancers. *Front. Cell Dev. Biol.* **10**, (2022).
56. Deng, C. *et al.* Pan-cancer analysis of CDKN2A alterations identifies a subset of gastric cancer with a cold tumor immune microenvironment. *Hum. Genomics* **18**, 55 (2024).
57. Wang, X. *et al.* Transcriptional analysis of the expression, prognostic value and immune infiltration activities of the COMMD protein family in hepatocellular carcinoma. *BMC Cancer* **21**, 1001 (2021).
58. Lau, D. K., Collin, J. P. & Mariadason, J. M. Clinical Developments and Challenges in Treating FGFR2-Driven Gastric Cancer. *Biomedicines* **12**, 1117 (2024).
59. Gordon, A., Johnston, E., Lau, D. K. & Starling, N. Targeting FGFR2 Positive Gastroesophageal Cancer: Current and Clinical Developments. *OncoTargets Ther.* **15**, 1183–1196 (2022).
60. Tsimafeyeu, I. & Raskin, G. Challenges of FGFR2 Testing in Gastric Cancer. *Oncol. Rev.* **17**, 11790 (2023).
61. Wahner, A. Anti-CEACAM5 ADC Could Represent the Next Targeted Therapy in Advanced Nonsquamous NSCLC. **1**, (2022).
62. Doherty, K. CEACAM5 Shows Great Potential as Therapeutic Target in NSCLC. **23**, (2022).
63. Hou, J. *et al.* ZC3H15 promotes gastric cancer progression by targeting the FBXW7/c-Myc pathway. *Cell Death Discov.* **8**, 32 (2022).
64. Anauate, A. C. *et al.* The Complex Network between MYC Oncogene and microRNAs in Gastric Cancer: An Overview. *Int. J. Mol. Sci.* **21**, 1782 (2020).
65. Frontiers | A Comprehensive Pan-Cancer Analysis of the Tumorigenic Role of Matrix Metallopeptidase 7 (MMP7) Across Human Cancers. https://www.frontiersin.org/journals/oncology/articles/10.3389/fonc.2022.916907/full.
66. Bornschein, J. *et al.* MMP2 and MMP7 at the invasive front of gastric cancer are not associated with mTOR expression. *Diagn. Pathol.* **10**, 212 (2015).

67. Cheng, Z., Zhang, D., Gong, B., Wang, P. & Liu, F. CD163 as a novel target gene of STAT3 is a potential therapeutic target for gastric cancer. *Oncotarget* **8**, 87244–87262 (2017).
68. Ashrafizadeh, M. *et al.* STAT3 Pathway in Gastric Cancer: Signaling, Therapeutic Targeting and Future Prospects. *Biology* **9**, 126 (2020).
69. Ye, Z. *et al.* Reprimo (RPRM) as a Potential Preventive and Therapeutic Target for Radiation-Induced Brain Injury via Multiple Mechanisms. *Int. J. Mol. Sci.* **24**, 17055 (2023).
70. Lee, T.-A. *et al.* Post-translational Modification of PD-1: Potential Targets for Cancer Immunotherapy. *Cancer Res.* **84**, 800–807 (2024).
71. Wang, R. *et al.* Emerging therapeutic frontiers in cancer: insights into posttranslational modifications of PD-1/PD-L1 and regulatory pathways. *Exp. Hematol. Oncol.* **13**, 46 (2024).
72. Zhang, J. *et al.* Alpha-fetoprotein predicts the treatment efficacy of immune checkpoint inhibitors for gastric cancer patients. *BMC Cancer* **24**, 266 (2024).
73. Augoff, K., Hryniewicz-Jankowska, A., Tabola, R. & Stach, K. MMP9: A Tough Target for Targeted Therapy for Cancer. *Cancers* **14**, 1847 (2022).
74. Zhou, Z., Xu, S., Jiang, L., Tan, Z. & Wang, J. A Systematic Pan-Cancer Analysis of CASP3 as a Potential Target for Immunotherapy. *Front. Mol. Biosci.* **9**, (2022).
75. Thuss-Patience, P. *et al.* Ramucirumab, Avelumab, and Paclitaxel as Second-Line Treatment in Esophagogastric Adenocarcinoma: The Phase 2 RAP (AIO-STO-0218) Nonrandomized Controlled Trial. *JAMA Netw. Open* **7**, e2352830 (2024).
76. Kanogawa, N. *et al.* Use of ramucirumab for various treatment lines in real-world practice of patients with advanced hepatocellular carcinoma. *BMC Gastroenterol.* **23**, 70 (2023).
77. Shitara, K. *et al.* Trastuzumab Deruxtecan in Previously Treated HER2-Positive Gastric Cancer. *N. Engl. J. Med.* **382**, 2419–2430 (2020).
78. Kang, Y.-K. *et al.* Rivoceranib, a VEGFR-2 inhibitor, monotherapy in previously treated patients with advanced or metastatic gastric or gastroesophageal junction cancer (ANGEL study): an international, randomized, placebo-controlled, phase 3 trial. *Gastric Cancer* **27**, 375–386 (2024).
79. Li, H., Shen, M. & Wang, S. Current therapies and progress in the treatment of advanced gastric cancer. *Front. Oncol.* **14**, (2024).
80. Zhiping H. *et al.* Evidence-based Evaluation on Tegafur-gimeracil-oteracil Potassium in the Treatment of Gastric Cancer. *Chin. J. Mod. Appl. Pharm.* **37**, 2644–2648 (2020).
81. Zhou, K. *et al.* Maintenance Therapy with Tegafur-gimeracil-oteracil After First-line Chemotherapy in Stage IV Gastric Cancer: A Retrospective Study. *Am. J. Clin. Exp. Med.* **9**, 168–173 (2021).
82. Dos Santos, M. *et al.* Perioperative treatment in resectable gastric cancer with spartalizumab in combination with fluorouracil, leucovorin, oxaliplatin and docetaxel (FLOT): a phase II study (GASPAR). *BMC Cancer* **22**, 537 (2022).
83. Gervaso, L. *et al.* Immunotherapy in the neoadjuvant treatment of gastrointestinal tumors: is the time ripe? *J. Immunother. Cancer* **12**, e008027 (2024).
84. Chinen, T., Sasabuchi, Y., Matsui, H., Yamaguchi, H. & Yasunaga, H. Oxaliplatin- versus cisplatin-based regimens for elderly individuals with advanced gastric cancer: a retrospective cohort study. *BMC Cancer* **22**, 460 (2022).
85. Aldossary, S. A. Review on Pharmacology of Cisplatin: Clinical Use, Toxicity and Mechanism of Resistance of Cisplatin. *Biomed. Pharmacol. J.* **12**, 7–15 (2019).
86. Giommoni, E. *et al.* Results of the observational prospective RealFLOT study. *BMC Cancer* **21**, 1086 (2021).
87. Shen, C. *et al.* Surgical treatment and prognosis of gastric neuroendocrine neoplasms: a single-center experience. *BMC Gastroenterol.* **16**, 111 (2016).

88. Hinnen, D. Glucagon-Like Peptide 1 Receptor Agonists for Type 2 Diabetes. *Diabetes Spectr.* **30**, 202–210 (2017).
89. Nishida, N., Sakai, D. & Satoh, T. Treatment strategy for HER2-negative advanced gastric cancer: salvage-line strategy for advanced gastric cancer. *Int. J. Clin. Oncol.* (2024) doi:10.1007/s10147-024-02500-8.
90. Pous, A., Notario, L., Hierro, C., Layos, L. & Bugés, C. HER2-Positive Gastric Cancer: The Role of Immunotherapy and Novel Therapeutic Strategies. *Int. J. Mol. Sci.* **24**, 11403 (2023).
91. Arai, H. *et al.* Fluoropyrimidine with or without platinum as first-line chemotherapy in patients with advanced gastric cancer and severe peritoneal metastasis: a multicenter retrospective study. *BMC Cancer* **19**, 652 (2019).
92. Lee, I.-S. *et al.* A liquid biopsy signature predicts treatment response to fluoropyrimidine plus platinum therapy in patients with metastatic or unresectable gastric cancer: implications for precision oncology. *Mol. Cancer* **21**, 9 (2022).
93. Arias-Martinez, A. *et al.* Is there a preferred platinum and fluoropyrimidine regimen for advanced HER2-negative esophagogastric adenocarcinoma? Insights from 1293 patients in AGAMENON–SEOM registry. *Clin. Transl. Oncol.* **26**, 1674–1686 (2024).
94. Joshi, S. S. & Badgwell, B. D. Current treatment and recent progress in gastric cancer. *CA. Cancer J. Clin.* **71**, 264–279 (2021).
95. Lv, H.-F. *et al.* Efficacy and safety of docetaxel plus S-1-based therapy in gastric cancer: a quantitative evidence synthesis of randomized controlled trials. *Front. Pharmacol.* **14**, (2024).
96. Leucovorin. https://go.drugbank.com/drugs/DB00650.
97. Gibson, M. K., Holcroft, C. A., Kvols, L. K. & Haller, D. Phase II Study of 5-fluorouracil, Doxorubicin, and Mitomycin C for Metastatic Small Bowel Adenocarcinoma. *The Oncologist* **10**, 132–137 (2005).
98. Lim, S. H. *et al.* Real-world outcomes of third-line immune checkpoint inhibitors versus irinotecan-based chemotherapy in patients with advanced gastric cancer: a Korean, multicenter study (KCSG ST22-06). *BMC Cancer* **24**, 252 (2024).
99. Chun, J. H. *et al.* Weekly Irinotecan in Patients with Metastatic Gastric Cancer Failing Cisplatin-based Chemotherapy. *Jpn. J. Clin. Oncol.* **34**, 8–13 (2004).
100. Roelofs, E. J. M. *et al.* Phase II study of sequential high-dose methotrexate (MTX) and 5-fluorouracil (F) alternated with epirubicin (E) and cisplatin (P) [FEMTX-P] in advanced gastric cancer. *Ann. Oncol.* **4**, 426–428 (1993).
101. De Vita, F. *et al.* Neo-adjuvant and adjuvant chemotherapy of gastric cancer. *Ann. Oncol.* **18**, vi120–vi123 (2007).
102. Jeong, S.-H. *et al.* Efficacy of S-1 or Capecitabine Plus Oxaliplatin Adjuvant Chemotherapy for Stage II or III Gastric Cancer after Curative Gastrectomy: A Systematic Review and Meta-Analysis. *Cancers* **14**, 3940 (2022).
103. Lee, S. *et al.* Real-world efficacy and safety of capecitabine with oxaliplatin in patients with advanced adenocarcinoma of the ampulla of Vater. *BMC Cancer* **24**, 634 (2024).
104. Wu, Y. *et al.* Hyperthermic intraperitoneal chemotherapy for patients with gastric cancer based on laboratory tests is safe: a single Chinese center analysis. *BMC Surg.* **22**, 342 (2022).
105. Rau, B. *et al.* Effect of Hyperthermic Intraperitoneal Chemotherapy on Cytoreductive Surgery in Gastric Cancer With Synchronous Peritoneal Metastases: The Phase III GASTRIPEC-I Trial. *J. Clin. Oncol.* (2023) doi:10.1200/JCO.22.02867.
106. Takeyoshi, I. *et al.* A Phase II Study of Weekly Paclitaxel and Doxifluridine Combination Chemotherapy for Advanced/Recurrent Gastric Cancer. *ANTICANCER Res.* (2011).
107. Ikeda, N. *et al.* A Phase II Study of Doxifluridine in Elderly Patients with Advanced Gastric Cancer: The Japan Clinical Oncology Group Study (JCOG 9410). *Jpn. J. Clin. Oncol.* **32**, 90–94 (2002).

108. Adashek, J. J., Arroyo-Martinez, Y., Menta, A. K., Kurzrock, R. & Kato, S. Therapeutic Implications of Epidermal Growth Factor Receptor (EGFR) in the Treatment of Metastatic Gastric/GEJ Cancer. *Front. Oncol.* **10**, (2020).
109. Lee, S.-H. *et al.* Combination chemotherapy with epirubicin, docetaxel and cisplatin (EDP) in metastatic or recurrent, unresectable gastric cancer. *Br. J. Cancer* **91**, 18–22 (2004).
110. Yuan, P. *et al.* Effect of Epirubicin Plus Paclitaxel vs Epirubicin and Cyclophosphamide Followed by Paclitaxel on Disease-Free Survival Among Patients With Operable ERBB2-Negative and Lymph Node–Positive Breast Cancer: A Randomized Clinical Trial. *JAMA Netw. Open* **6**, e230122 (2023).
111. Karmi, J. A. & Gibson, M. K. Zolbetuximab: An Investigational First-Line Treatment for CLDN18.2-positive, HER2-negative Gastric and Gastro-oesophageal Junction Cancer. *1* (2023).
112. Yen, C.-C. *et al.* Surgery alone, adjuvant tegafur/gimeracil/octeracil (S-1), or platinum-based chemotherapies for resectable gastric cancer: real-world experience and a propensity score matching analysis. *BMC Cancer* **21**, 796 (2021).
113. Ling, Q. *et al.* Optimal timing of surgery for gastric cancer after neoadjuvant chemotherapy: a systematic review and meta-analysis. *World J. Surg. Oncol.* **21**, 377 (2023).
114. Tanabe, S. Advances in Molecular Mechanisms of Gastrointestinal Tumors. *Cancers* **16**, 1603 (2024).
115. Wang, F.-Y. *et al.* Nanoparticle Polymeric Micellar Paclitaxel Versus Paclitaxel for Patients with Advanced Gastric Cancer. *J. Gastrointest. Cancer* (2024) doi:10.1007/s12029-024-01058-y.
116. Guan, W.-L., He, Y. & Xu, R.-H. Gastric cancer treatment: recent progress and future perspectives. *J. Hematol. Oncol.J Hematol Oncol* **16**, 57 (2023).
117. Hassan, M. S., Awasthi, N., Ponna, S. & von Holzen, U. Nab-Paclitaxel in the Treatment of Gastrointestinal Cancers—Improvements in Clinical Efficacy and Safety. *Biomedicines* **11**, 2000 (2023).
118. Shibata, C. *et al.* Comparison of CEA and CA19-9 as a predictive factor for recurrence after curative gastrectomy in gastric cancer. *BMC Surg.* **22**, 213 (2022).
119. Ma, X. *et al.* CA19-9 is a significant prognostic factor in stage III gastric cancer patients undergoing radical gastrectomy. *BMC Surg.* **24**, 31 (2024).
120. Xu, H.-M. *et al.* Helicobacter pylori Treatment and Gastric Cancer Risk Among Individuals With High Genetic Risk for Gastric Cancer. *JAMA Netw. Open* **7**, e2413708 (2024).
121. Zheng, Z., Lu, Z. & Song, Y. Long-term proton pump inhibitors use and its association with premalignant gastric lesions: a systematic review and meta-analysis. *Front. Pharmacol.* **14**, (2023).
122. Salari, N. *et al.* The global prevalence of sexual dysfunction in women with multiple sclerosis: a systematic review and meta-analysis. *Neurol. Sci.* **44**, 59–66 (2023).
123. Kiaei, A. A. *et al.* FPL: False Positive Loss. (2023).
124. Kiaei, A. A. *et al.* Active Identity Function as Activation Function. (2023).
125. Kiaei, A. A. *et al.* Diagnosing Alzheimer's Disease Levels Using Machine Learning and MRI: A Novel Approach. (2023).
126. Behrouzi, Y., Basiri, A., Pourgholi, R. & Kiaei, A. A. Fusion of medical images using Nabla operator; Objective evaluations and step-by-step statistical comparisons. *Plos One* **18**, e0284873 (2023).
127. Salari, N. *et al.* The effects of smoking on female sexual dysfunction: a systematic review and meta-analysis. *Arch. Womens Ment. Health* **25**, 1021–1027 (2022).
128. Salari, N. *et al.* Global prevalence of Duchenne and Becker muscular dystrophy: a systematic review and meta-analysis. *J. Orthop. Surg.* **17**, 96 (2022).
129. Jafari, H. *et al.* A full pipeline of diagnosis and prognosis the risk of chronic diseases using deep learning and Shapley values: The Ravansar county anthropometric cohort study. *PloS One* **17**, e0262701 (2022).
130. Kiaei, A. A. & Khotanlou, H. Segmentation of medical images using mean value guided contour. *Med. Image Anal.* **40**, 111–132 (2017).

131. Parichehreh, E. *et al.* Graph Attention Networks for Drug Combination Discovery: Targeting Pancreatic Cancer Genes with RAIN Protocol. *medRxiv* 2024–02 (2024).
132. Salari, N. *et al.* Executive protocol designed for new review study called: systematic review and artificial intelligence network meta-analysis (RAIN) with the first application for COVID-19. *Biol. Methods Protoc.* **8**, bpac038 (2023).
133. Safaei, D. *et al.* Systematic review and network meta-analysis of drug combinations suggested by machine learning on genes and proteins, with the aim of improving the effectiveness of Ipilimumab in treating Melanoma. *medRxiv* 2023–05 (2023).
134. Kiaei, A. A. *et al.* Recommending Drug Combinations Using Reinforcement Learning targeting Genes/proteins associated with Heterozygous Familial Hypercholesterolemia: A comprehensive Systematic Review and Net-work Meta-analysis. (2023).
135. Kiaei, A. A. *et al.* Recommending Drug Combinations using Reinforcement Learning to target Genes/proteins that cause Stroke: A comprehensive Systematic Review and Network Meta-analysis. (2023).
136. Boush, M., Kiaei, A. A. & Mahboubi, H. Trending Drugs Combination to Target Leukemia associated Proteins/Genes: using Graph Neural Networks under the RAIN Protocol. *medRxiv* 2023–08 (2023).
137. Boush, M. *et al.* Drug combinations proposed by machine learning on genes/proteins to improve the efficacy of Tecovirimat in the treatment of Monkeypox: A Systematic Review and Network Meta-analysis. *medRxiv* 2023–04 (2023).
138. Boush, M. *et al.* Recommending Drug Combinations using Reinforcement Learning to target Genes/proteins that cause Stroke: A comprehensive Systematic Review and Network Meta-analysis. *medRxiv* 2023–04 (2023).
139. Kiaei, A. *et al.* Identification of suitable drug combinations for treating COVID-19 using a novel machine learning approach: the RAIN Method. *Life* **12**, 1456 (2022).
140. Mohammadi, M., Salari, N., Far, A. H. & Kiaei, A. Executive protocol designed for new review study called: Systematic Review and Artificial Intelligence Network Meta-Analysis (RAIN) with the first application for COVID-19. (2021).
141. Yang, C., Xiao, C., Ma, F., Glass, L. & Sun, J. SafeDrug: Dual Molecular Graph Encoders for Recommending Effective and Safe Drug Combinations. Preprint at http://arxiv.org/abs/2105.02711 (2022).
142. Dashti, N., Kiaei, A. A., Boush, M., Gholami-Borujeni, B. & Nazari, A. AI-Enhanced RAIN Protocol: A Systematic Approach to Optimize Drug Combinations for Rectal Neoplasm Treatment. *bioRxiv* 2024–05 (2024).
143. Gu, J. *et al.* A model-agnostic framework to enhance knowledge graph-based drug combination prediction with drug–drug interaction data and supervised contrastive learning. *Brief. Bioinform.* **24**, bbad285 (2023).
144. Salari, N. *et al.* Global prevalence of vasovagal syncope: A systematic review and meta-analysis. *Glob. Epidemiol.* **7**, 100136 (2024).
145. Zhou, R. *et al.* NEDD: a network embedding based method for predicting drug-disease associations. *BMC Bioinformatics* **21**, 387 (2020).
146. Han, Y., Klinger, K., Rajpal, D. K., Zhu, C. & Teeple, E. Empowering the discovery of novel target-disease associations via machine learning approaches in the open targets platform. *BMC Bioinformatics* **23**, 232 (2022).
147. Shi, Y. *et al.* Long non-coding RNAs in drug resistance across the top five cancers: Update on their roles and mechanisms. *Heliyon* **10**, e27207 (2024).
148. T, H. *et al.* Preoperative docetaxel, cisplatin, and 5-fluorouracil for resectable locally advanced esophageal and esophagogastric junctional adenocarcinoma. *Esophagus Off. J. Jpn. Esophageal Soc.* **21**, (2024).

149. K, B. Y. *et al.* Comparison of the second-line treatment efficacy in advanced gastric cancer patients previously treated with taxane-based triplet chemotherapy: a Turkish Oncology Group Study. *Curr. Med. Res. Opin.* **40**, (2024).
150. Y, Z. *et al.* Personalized drug screening using patient-derived organoid and its clinical relevance in gastric cancer. *Cell Rep. Med.* **5**, (2024).
151. Mz, Q. *et al.* Tislelizumab plus chemotherapy versus placebo plus chemotherapy as first line treatment for advanced gastric or gastro-oesophageal junction adenocarcinoma: RATIONALE-305 randomised, double blind, phase 3 trial. *BMJ* **385**, (2024).
152. Zhao, S. *et al.* Phase I trial of apatinib and paclitaxel+oxaliplatin+5-FU/levoleucovorin for treatment-naïve advanced gastric cancer. *Cancer Sci.* **115**, 1611–1621 (2024).
153. G, C. *et al.* Circumventing drug resistance in gastric cancer: A spatial multi-omics exploration of chemo and immuno-therapeutic response dynamics. *Drug Resist. Updat. Rev. Comment. Antimicrob. Anticancer Chemother.* **74**, (2024).
154. D, M., R, M., D, M., A, M. & J, P.-S. Synchronous Gastric and Rectal Adenocarcinoma With Complete Response After Total Neoadjuvant Therapy: A Case Report. *Cureus* **16**, (2024).
155. P, T., A, K., J, Y., S, A. & S, K. The Debut Signal of Bone Metastasis and Stealthy Gastric Cancer Unmasked in a Young Male: A Case Report. *Cureus* **16**, (2024).
156. F, N., Q, G. & Pm, H. Nivolumab-Induced Cytokine Release Syndrome: A Case Report and Literature Review. *Am. J. Case Rep.* **25**, (2024).
157. Mineur, L. *et al.* NESC Multicenter Phase II Trial in the Preoperative Treatment of Gastric Adenocarcinoma with Chemotherapy (Docetaxel-Cisplatin-5FU+Lenograstim) Followed by Chemoradiation Based 5FU and Oxaliplatin and Surgery. *Cancer Res. Treat.* **56**, 580–589 (2024).
158. Li, N. *et al.* A randomized phase 2 study of HLX22 plus trastuzumab biosimilar HLX02 and XELOX as first-line therapy for HER2-positive advanced gastric cancer. *Med N. Y. N* S2666-6340(24)00250–2 (2024) doi:10.1016/j.medj.2024.06.004.
159. Wakatsuki, T. *et al.* Exploratory analysis of serum HER2 extracellular domain for HER2 positive gastric cancer treated with SOX plus trastuzumab. *Int. J. Clin. Oncol.* **29**, 801–812 (2024).
160. Yanagisawa, N. *et al.* Two cases of gastric cancer with elevated serum levels of KL-6. *Surg. Case Rep.* **10**, 82 (2024).
161. Hirase, Y. *et al.* Successful subtotal gastrectomy and hepatectomy for HER2-positive gastric cancer with liver metastasis after trastuzumab-based chemotherapy: a case report. *Surg. Case Rep.* **10**, 51 (2024).
162. Zhou, J.-H. *et al.* Inetetamab combined with tegafur as second-line treatment for human epidermal growth factor receptor-2-positive gastric cancer: A case report. *World J. Clin. Cases* **12**, 820–827 (2024).
163. Yamane, H. *et al.* Long-Term Complete Response to Trastuzumab Deruxtecan in a Case of Unresectable Gastric Cancer. *Case Rep. Oncol.* **17**, 463–470 (2024).
164. Fuchino, M. *et al.* [A case of mixed neuroendocrine-non-neuroendocrine neoplasm (MiNEN) diagnosed by re-biopsy of the enlarged primary tumor during chemotherapy for gastric adenocarcinoma]. *Nihon Shokakibyo Gakkai Zasshi Jpn. J. Gastro-Enterol.* **121**, 55–62 (2024).
165. Liu, Y. *et al.* Current Progress on Predictive Biomarkers for Response to Immune Checkpoint Inhibitors in Gastric Cancer: How to Maximize the Immunotherapeutic Benefit? *Cancers* **15**, 2273 (2023).
166. Yoon, J., Kim, T.-Y. & Oh, D.-Y. Recent Progress in Immunotherapy for Gastric Cancer. *J. Gastric Cancer* **23**, 207–223 (2023).
167. Shirasaki, Y. *et al.* [A Case of HER2-Positive Unresectable Advanced Gastric Cancer Showing Long-Term Complete Response Treated by Only Chemotherapy]. *Gan To Kagaku Ryoho* **50**, 1887–1888 (2023).

168. Shimizu, K. *et al.* [A Case of HER2-Positive Advanced Gastric Cancer with Liver Metastasis Treated by Laparoscopic Surgery after Chemotherapy and Achieved Pathological Complete Response]. *Gan To Kagaku Ryoho* **50**, 1801–1803 (2023).
169. Li, S. *et al.* Multicenter phase I dose escalation and expansion study of pyrotinib in combination with camrelizumab and chemotherapy as first-line treatment for HER2-positive advanced gastric and gastroesophageal junction adenocarcinoma. *EClinicalMedicine* **66**, 102314 (2023).
170. Iwamuro, M., Tanaka, T., Kagawa, S., Inoo, S. & Otsuka, M. Collagenous Colitis in a Patient With Gastric Cancer Who Underwent Chemotherapy. *Cureus* **15**, e39466 (2023).
171. Huo, B., Lin, L., Zhao, L., Yu, R. & Yang, J. First reported case of ANCA-associated vasculitis induced by oxaliplatin, capecitabine, and trastuzumab. *Ren. Fail.* **45**, 2282710 (2023).
172. Hofheinz, R.-D. *et al.* FLOT Versus FLOT/Trastuzumab/Pertuzumab Perioperative Therapy of Human Epidermal Growth Factor Receptor 2-Positive Resectable Esophagogastric Adenocarcinoma: A Randomized Phase II Trial of the AIO EGA Study Group. *J. Clin. Oncol. Off. J. Am. Soc. Clin. Oncol.* **40**, 3750–3761 (2022).
173. Aktürk Esen, S. *et al.* First-line treatment of patients with HER2-positive metastatic gastric and gastroesophageal junction cancer. *Bosn. J. Basic Med. Sci.* **22**, 818–825 (2022).
174. Nishie, N. *et al.* Successful open radical gastrectomy for locally advanced or metastatic gastric cancer patients who suffered from coronavirus disease 2019 during preoperative chemotherapy: a report of three cases. *Surg. Case Rep.* **8**, 124 (2022).
175. Zhang, Q. *et al.* Temporal heterogeneity of HER2 expression in metastatic gastric cancer: a case report. *World J. Surg. Oncol.* **20**, 157 (2022).
176. Schade, S. *et al.* Cure Is Possible: Extensively Metastatic HER2-Positive Gastric Carcinoma with 5 years of Complete Remission after Therapy with the FLOT Regimen and Trastuzumab. *Case Rep. Gastroenterol.* **16**, 80–88 (2022).
177. Ladak, I., Preti, B., Dias, B. & Breadner, D. A. Raltitrexed as a substitute for capecitabine in metastatic gastric cancer: a case report and literature review. *Ann. Transl. Med.* **10**, 1285 (2022).
178. Wu, X.-Q., Ge, Y.-P., Gong, X.-L., Liu, Y. & Bai, C.-M. [Advances in Treatment of Human Epidermal Growth Factor Receptor 2-Positve Gastric Cancer]. *Zhongguo Yi Xue Ke Xue Yuan Xue Bao* **44**, 899–905 (2022).
179. Mori, Y. *et al.* Phase II Prospective Study of Trastuzumab in Combination with S-1 and Oxaliplatin (SOX100) Therapy for HER2-Positive Advanced Gastric Cancer. *J. Gastrointest. Cancer* **53**, 930–938 (2022).
180. Hofheinz, R.-D. *et al.* Trastuzumab in combination with 5-fluorouracil, leucovorin, oxaliplatin and docetaxel as perioperative treatment for patients with human epidermal growth factor receptor 2-positive locally advanced esophagogastric adenocarcinoma: A phase II trial of the Arbeitsgemeinschaft Internistische Onkologie Gastric Cancer Study Group. *Int. J. Cancer* **149**, 1322–1331 (2021).
181. Grieb, B. C. & Agarwal, R. HER2-Directed Therapy in Advanced Gastric and Gastroesophageal Adenocarcinoma: Triumphs and Troubles. *Curr. Treat. Options Oncol.* **22**, 88 (2021).
182. Horinouchi, T. *et al.* Human Epidermal Growth Factor Receptor 2-positive Primary Adenocarcinoma in the Cervical Oesophagus: A Case Report. *Vivo Athens Greece* **35**, 2297–2303 (2021).
183. Jimenez-Fonseca, P. *et al.* External validity of clinical trials with diverse trastuzumab-based chemotherapy regimens in advanced gastroesophageal adenocarcinoma: data from the AGAMENON-SEOM registry. *Ther. Adv. Med. Oncol.* **13**, 17588359211019672 (2021).
184. Yamashita, K., Hosoda, K., Niihara, M. & Hiki, N. History and emerging trends in chemotherapy for gastric cancer. *Ann. Gastroenterol. Surg.* **5**, 446–456 (2021).

185. Al-Batran, S.-E. *et al.* Clinical Practice Observation of Trastuzumab in Patients with Human Epidermal Growth Receptor 2-Positive Metastatic Adenocarcinoma of the Stomach or Gastroesophageal Junction. *The Oncologist* **25**, e1181–e1187 (2020).
186. Park, H. *et al.* FOLFIRINOX for the Treatment of Advanced Gastroesophageal Cancers: A Phase 2 Nonrandomized Clinical Trial. *JAMA Oncol.* **6**, 1231–1240 (2020).
187. Janjigian, Y. Y. *et al.* First-line pembrolizumab and trastuzumab in HER2-positive oesophageal, gastric, or gastro-oesophageal junction cancer: an open-label, single-arm, phase 2 trial. *Lancet Oncol.* **21**, 821–831 (2020).
188. Tintelnot, J. *et al.* Ipilimumab or FOLFOX with Nivolumab and Trastuzumab in previously untreated HER2-positive locally advanced or metastatic EsophagoGastric Adenocarcinoma - the randomized phase 2 INTEGA trial (AIO STO 0217). *BMC Cancer* **20**, 503 (2020).
189. Catenacci, D. V. T. *et al.* Evaluation of the Association of Perioperative UGT1A1 Genotype-Dosed gFOLFIRINOX With Margin-Negative Resection Rates and Pathologic Response Grades Among Patients With Locally Advanced Gastroesophageal Adenocarcinoma: A Phase 2 Clinical Trial. *JAMA Netw. Open* **3**, e1921290 (2020).
190. Gutting, T. *et al.* Complete Remission of Metastatic HER2+ Oesophagogastric Junctional Adenocarcinoma under long-term Trastuzumab Treatment. *J. Gastrointest. Liver Dis. JGLD* **28**, 503–507 (2019).